\pdfoutput=1
\documentclass[11pt,twoside,a4paper,cmspaper,final,collab]{cms-tdr}
\def\svnVersion{571d792}\def\svnDate{2025/01/06}\def\cmsCernNoTag{CERN-EP-2024-088}\def\cmsCernDate{\today}\def\cmsMessage{Published in Physics Letters B as \href{http://dx.doi.org/10.1016/j.physletb.2024.139067}{\doi{10.1016/j.physletb.2024.139067}.}}
\begin{document}\cmsNoteHeader{HIG-20-002}

\newlength\cmsFigWidth
\newlength\cmsTabSkip\setlength{\cmsTabSkip}{1ex}
\ifthenelse{\boolean{cms@external}}{\setlength\cmsFigWidth{0.98\columnwidth}}{\setlength\cmsFigWidth{0.6\textwidth}}
\ifthenelse{\boolean{cms@external}}{\providecommand{\cmsLeft}{upper\xspace}}{\providecommand{\cmsLeft}{left\xspace}}
\ifthenelse{\boolean{cms@external}}{\providecommand{\cmsRight}{lower\xspace}}{\providecommand{\cmsRight}{right\xspace}}

\ifthenelse{\boolean{cms@external}}{\providecommand{\cmsTable}[1]{#1}}{\providecommand{\cmsTable}[1]{\resizebox{\textwidth}{!}{#1}}}
\newcommand{\mH}{\ensuremath{m_{\PH}}\xspace}
\newcommand{\mgg}{\ensuremath{m_{\PGg\PGg}}\xspace}
\newcommand{\ptga}{\ensuremath{p_\mathrm{T}^{\PGg1}}\xspace}
\newcommand{\ptgb}{\ensuremath{p_\mathrm{T}^{\PGg2}}\xspace}

\cmsNoteHeader{HIG-20-002}
\title{Search for a standard model-like Higgs boson in the mass range between 70 and \texorpdfstring{110\GeV}{110 GeV} in the diphoton final state in proton-proton collisions at \texorpdfstring{$\sqrt{s}=13\TeV$}{sqrt(s) = 13 TeV}}

\date{\today}

\abstract{The results of a search for a standard model-like Higgs boson decaying into two photons in the mass range between 70 and 110\GeV are presented.  The analysis uses the data set collected by the CMS experiment in proton-proton collisions at $\sqrt{s}=13\TeV$ corresponding to integrated luminosities of 36.3\fbinv, 41.5\fbinv and 54.4\fbinv during the 2016, 2017, and 2018 LHC running periods, respectively.  No significant excess over the background expectation is observed and 95\% confidence level upper limits are set on the product of the cross section and branching fraction for decays of an additional Higgs boson into two photons.  The maximum deviation with respect to the background is seen for a mass hypothesis of 95.4\GeV with a local (global) significance of 2.9 (1.3) standard deviations.  The observed upper limit ranges from 15 to 73\unit{fb}.}

\hypersetup{
pdfauthor={CMS Collaboration},
pdftitle={Search for a standard model-like Higgs boson in the mass range between 70 and 110 GeV  in the diphoton final state in proton-proton collisions at sqrt(s) = 13 TeV},
pdfsubject={CMS},
pdfkeywords={CMS,  Higgs, diphoton}}

\maketitle

\section{Introduction}
\label{sec:intro}
In the standard model (SM) of particle physics~\cite{SMHiggsO,SMHiggsT,SMHiggsTh}, 
the masses of fundamental particles arise from the spontaneous breaking of  electroweak symmetry, 
in the manner of the Brout--Englert--Higgs mechanism~\cite{HiggsExpO,HiggsExpT,HiggsExpTh,PhysRevLett.13.585,
PhysRev.145.1156,PhysRev.155.1554}. In its minimal version, 
this mechanism introduces a doublet of complex scalar fields. 
After symmetry breaking, only one scalar field
remains, giving rise to the physical 
Higgs boson, that  
is experimentally observable.  In 2012,  
the ATLAS~\cite{Aad:2012tfa} and
CMS~\cite{Chatrchyan:2012xdj,Chatrchyan:2013lba} collaborations observed a new boson with
a mass of approximately 125\GeV whose properties are at present compatible, within uncertainties~\cite{ATLASNature,CMSNature}, with those of the
SM Higgs boson.
The analyses of data in the diphoton final state leading to this discovery probed an invariant mass range extending
from 110
to 150\GeV.
 
{\tolerance=800 Several models for physics beyond the SM (BSM)  
typically contain multiple scalars with one of them being 
compatible with the observed 125\GeV boson.
The extended parameter space of
several BSM models,
for example, generalized
models containing two Higgs doublets 
(2HDM)~\cite{Celis_2013,Coleppa:2013dya,Chang:2013ona,Bernon:2015wef,Cacciapaglia2016}, the next-to-minimal supersymmetric model (NMSSM)~\cite{Fayet:1974pd,Barbieri:1982eh,Dine:1981rt,Nilles:1982dy,Frere:1983ag,Derendinger:1983bz,PhysRevD.39.844,Drees:1988fc,Ellwanger:1993xa,Ellwanger:1995ru,Ellwanger:1996gw,Elliott:1994ht,King:1995vk,Franke:1995tc,Maniatis:2009re,Ellwanger:2009dp,1674-1137-38-7-073101,Ellwanger:2015uaz,Guchait:2016pes,Cao:2016uwt,Tao_2018}, and Georgi--Machacek models~\cite{GEORGI1985463,PhysRevD.92.075011,PhysRevD.103.095010,Wang_2022}, gives rise to rich and interesting phenomenology, including the presence of
additional 
scalars, some of which could have masses below 125\GeV.
Such models provide good motivation for
extending searches for
additional Higgs bosons 
with masses \mH
as far
below 110\GeV as possible, where \PH refers to an additional Higgs boson
that could also be ``SM-like", meaning that the relative contributions of the production and decay processes are
similar to those of the SM.
\par}

  The \HGG decay channel provides a clean final-state topology
that allows the mass of
a Higgs boson to be reconstructed with high precision.  The primary production mechanism
for SM-like Higgs bosons in proton-proton ($\Pp\Pp$) collisions
at
the CERN LHC
is gluon fusion ($\Pg\Pg\PH$), with smaller contributions from vector boson fusion (VBF) and production in association with a \PW or \PZ boson ($\mathrm{V}\PH$), or with a \ttbar pair ($\ttbar\PH$). The dominant sources of background are
irreducible direct diphoton production, and the reducible \PPTOGAMJET\ and $\Pp\Pp\to\text{jet}+\text{jet}$ processes, where the jets are misidentified as isolated photons. An additional source of reducible background relevant for the search range  $\mH<110\GeV$ is
Drell--Yan (DY) dielectron production, where electrons might be misidentified as isolated photons.

The CERN LEP collaborations~\cite{Barate:2003sz}, in the context of the search for the SM Higgs boson, extensively explored the mass range below 110\GeV in the $\mathrm{V}\PH$ production modes,
in the \bbbar and \TT
channels. Several of the BSM models mentioned above permit a significant region of parameter space for enhanced decay rates  in the diphoton channel compatible with the LEP constraints.  The ``low-mass" searches in the diphoton decay channel by 
ATLAS, performed in the diphoton invariant mass (\mgg) ranges of $65<\mgg<110\GeV$ at a center-of-mass energy of 8\TeV~\cite{PhysRevLett.113.171801} and $10<\mgg<70\GeV$ at a center-of-mass energy of 13\TeV~\cite{Aad2023}, the latter targeting boosted resonances, 
found no significant excess of events.
A CMS search~\cite{HIG-17-013} performed at center-of-mass energies of 8\,(13)\TeV in the mass ranges of $80\,(70)<\mgg<110\GeV$ reported an excess with respect to the 
SM background expectation, maximal for a mass hypothesis of 95.3\GeV, with a local (global) significance of 2.8\,(1.3) standard deviations. 

This
Letter presents an update of the CMS search in the diphoton channel for an
additional
Higgs boson in the invariant mass range $70<\mgg<110\GeV$; the natural width of this additional Higgs boson is small compared to the detector resolution.
The search is performed with a data set collected in 2016, 2017, and 2018
with the
CMS detector at the
LHC, corresponding to integrated luminosities of 36.3\fbinv, 41.5\fbinv, and 54.4\fbinv, respectively, at a center-of-mass energy of 13\TeV. As part of this study, the data set of 2016, used in Ref.~\cite{HIG-17-013}, has been reanalyzed with an improved detector calibration.  The tabulated results are provided as HepData records~\cite{hepdata}.

The analysis is based on a search for a localized excess
in the diphoton invariant mass spectrum.  We follow the method used in Ref.~\cite{HIG-17-013}, which is an extended version of the method developed for the observation and the measurement of the properties of the 125\GeV
boson~\cite{HIG-13-001,
CMS-PAS-HIG-16-040}.  The principal challenges associated with a search in the diphoton decay channel in the explored mass range are the ability to trigger on events while maintaining acceptable rates, and the background from
\PZ bosons decaying to electron pairs that, through misidentification, could appear as two isolated photons.  With respect to Ref.~\cite{HIG-17-013}, additional requirements have been introduced to further discriminate against these surviving electron pairs.  To achieve the best possible sensitivity, the events are separated into classes, including, for the data sets collected in 2017 and 2018, a class requiring the presence of additional jets targeting the VBF production mode.  
Multivariate analysis (MVA) techniques are used both for photon identification and event classification, and
the signal is extracted from the background using a parametric fit to the diphoton mass spectrum in all event classes. 

\section{The CMS detector}
\label{sec:det}

A detailed description of the CMS detector, together with a definition of the coordinate system used and the relevant kinematic variables, can be found in
Ref.~\cite{Chatrchyan:2008zzk}.  The central feature of the CMS
apparatus is a superconducting solenoid of 6\unit{m} internal diameter,  providing a magnetic field of 3.8\unit{T}. Within the
solenoid volume are a silicon pixel and strip tracker, a lead tungstate crystal electromagnetic calorimeter (ECAL), and a brass and scintillator hadron calorimeter (HCAL),  each composed of a barrel and two endcap sections. Forward calorimeters extend the pseudorapidity ($\eta$)
coverage provided by the barrel and endcap detectors. Muons are reconstructed in gas-ionization chambers embedded in the steel flux-return yoke outside the solenoid.
The ECAL, surrounding the tracker volume, consists of
75\,848 lead tungstate crystals, which provide coverage in
$\abs{ \eta }< 1.48$ in a barrel region (EB) and $1.48 <\abs{ \eta } < 3.0$ in two endcap regions (EE).
Preshower detectors consisting of two planes of silicon sensors interleaved with a total of $3 X_0$ of lead are located in front of each EE detector.
In the EB, an energy resolution of about 1\% is achieved for unconverted or late-converting photons in the tens of \GeV energy range.  The energy resolution of the remaining barrel photons is about 1.3\% up to $\abs{\eta}=1$, changing to about 2.5\% at $\abs{\eta}=1.4$. In the
EE, the energy resolution is about 2.5\% for unconverted or late-converting photons, and between 3 and 4\% for the other ones~\cite{CMS:EGM-14-001}. The diphoton mass resolution, as measured in \HGG decays, is typically in the 1--2\% range, depending on the measurement of the photon energies in the ECAL and the topology of the photons in the event~\cite{CMS:2020xrn}. 

The particle-flow algorithm~\cite{CMS-PRF-14-001} aims to reconstruct and identify each individual particle in an event, with an optimized combination of information from the various elements of the CMS detector. The energy of photons is obtained from the ECAL measurement. The energy of electrons is determined from a combination of the electron momentum at the primary interaction vertex as determined by the tracker, the energy of the corresponding ECAL cluster, and the energy sum of all bremsstrahlung photons spatially compatible with originating from the electron track. The energy of muons is obtained from the curvature of the corresponding track. The energy of charged hadrons is determined from a combination of their momentum measured in the tracker and the matching ECAL and HCAL energy deposits, corrected for the response function of the calorimeters to hadronic showers. Finally, the energy of neutral hadrons is obtained from the corresponding corrected ECAL and HCAL energies. 

\section{Event reconstruction, selection and categorization}
\subsection{Simulation and Trigger}
\label{sec:triggerSim}

Monte Carlo
(MC) simulations are used to produce
SM-like Higgs boson events from
all production processes
($\Pg\Pg\PH$, VBF, $\mathrm{V}\PH$, and $\ttbar\PH$), with Higgs boson masses ranging from 70 to 110\GeV in steps of 5\GeV.
These events are the input to the signal modeling procedure,
representing a new resonance decaying to two photons. They are generated at next-to-leading order (NLO) in perturbative quantum chromodynamics (QCD) using \MGvATNLO 2.4.2~\cite{Alwall:2014hca} with FxFx merging~\cite{FXFX} 
and the 
NNPDF3.0~\cite{Ball:2014uwa} 
(NNPDF3.1~\cite{NNPDF31}) set of parton distribution functions (PDFs) to simulate the 2016 (2017 and 2018) data.
Events produced via the $\Pg\Pg\PH$ process are weighted as a function of the Higgs boson \pt
and the number of jets in the event to match the prediction from the NNLOPS program~\cite{Hamilton:2013fea}.  The parton-level samples are interfaced with \PYTHIA 8.226 (8.230)~\cite{Sjostrand:2014zea} for parton showering and hadronization in the simulation of the 2016 (2017 and 2018) data, with the CUETP8M1~\cite{Khachatryan:2015pea} 
(CP5~\cite{Sirunyan:2019dfx})
tune parameter set used for the underlying event activity.
The cross sections and branching
fractions recommended by the LHC Higgs cross section working group for a center-of-mass energy of 13\TeV~\cite{deFlorian:2016spz}
are used for modeling the signal.
After the generation step, the events are
processed by the full CMS detector simulation based on \GEANTfour~\cite{GEANT}. Multiple
$\Pp\Pp$ interactions in each bunch crossing (pileup) are simulated.  The
events are weighted to reproduce the distribution of
the number of interactions observed in data, the average value of which was 23 in 2016 and 32 for both 2017 and 2018.

Events corresponding to the SM background processes mentioned in Section~\ref{sec:intro} are simulated using various generators. The diphoton background is modeled with the \SHERPA 2.2.4
\cite{Gleisberg:2008ta} generator; it includes tree-level diagrams with up to 3 additional jets,
as well as 
box diagrams at
leading order (LO).  Multijet and \GAMJET\ backgrounds are modeled at LO with \PYTHIA 8.226
(\PYTHIA 8.230) for the simulation of the 2016 (2017 and 2018) data,  
with a
filter~\cite{HIG-13-001,CMS-PAS-HIG-16-040} applied at generator level in order to
enhance the production of jets with a large fraction of electromagnetic
energy.  DY events are simulated at NLO with
\MGvATNLO 2.4.2.  All background events are generated using the same PDF sets and simulated under the same conditions as the SM-like Higgs boson events described above.
The simulated background samples are used in the calculation of corrections and efficiencies, training and validation of the multivariate boosted decision trees (BDTs) used in the analysis, and in  estimations of systematic
uncertainties.  In particular, the DY events are used for modeling the background contribution from dielectron decays of the \PZ boson.  As in Refs.~\cite{HIG-13-001,CMS-PAS-HIG-16-040}, the background estimation is based on data.

Events of interest are selected using a two-tiered trigger system. The first level,
composed of custom hardware processors, uses information from the calorimeters and muon detectors to select events at a rate of around 100\unit{kHz} within a fixed latency of about 4\unit{\mus}~\cite{CMS:2020cmk}. The second level, known as the high-level trigger (HLT), consists of a farm of processors running a version of the full event reconstruction software optimized for fast processing, and reduces the event rate to around 1\unit{kHz} before data storage~\cite{Khachatryan:2016bia}. 

For this analysis,
diphoton HLT paths with asymmetric photon transverse momentum
(\pt) thresholds, 30 and 18\GeV, are used for the data of all three years. For the 2016 data, two paths are used: one path has nearly identical requirements to those used in Ref.~\cite{CMS-PAS-HIG-16-040}, except that only events with both photon candidates in the EB are selected. This path requires 
each of the photon candidates to satisfy criteria on the 
maximum value of the ratio of its energy in the HCAL to its energy in the ECAL (H/E), and on 
either 
the compatibility of its shower shape 
with that of a photon, or on the maximum value of its isolation energy~\cite{Sirunyan:2020ycc}.  The other path selects events with photon candidates from any part of the ECAL, but they must satisfy 
more stringent shower shape requirements, as well as the requirements on both isolation energy and H/E.  Both paths impose a veto on the presence of hits compatible with the photon direction in the silicon pixel detector, and
require that the  invariant mass of the two photon candidates 
be greater than 55\GeV.

For the other two years of data taking, only one path, a relaxed variant of the latter path above, is used: for the 2017 data, at least one of the two photon candidates must satisfy the more stringent shower shape requirements, and for the 2018 data, the more stringent shower shape requirements
are applied to all photon candidates in the EE, enabling the removal of the silicon pixel detector hit veto
and invariant mass requirements.  However, the 2018 version of the path was not introduced until after the start of that year's data taking, resulting in a reduction of the data set by approximately 5\fbinv.

These HLT requirements impose a diphoton invariant mass value of $\mgg=65\GeV$ as the lower limit of the background modeling range; this value lies above the portion of the offline diphoton spectrum that is distorted due to turn-on effects from the HLT criteria.  Furthermore, in order to avoid edge effects in the signal modeling, the search range is limited to values lying between the minimal and maximal simulated Higgs boson masses:  $70<\mgg<110\GeV$.  The trigger efficiency is measured using $\PZ\to\EE$ events and the ``tag-and-probe" technique~\cite{tagNprobe}, except for the pixel detector hit veto requirement implemented for the triggering of the 2016 and 2017 data; the efficiency of this veto requirement is measured using diphoton events that have passed a diphoton trigger that does not require a pixel hit veto.  The efficiencies are applied to the simulated SM Higgs boson events as a correction, and the associated statistical and systematic uncertainties are propagated to the expected signal yields.  A gradual shift in the timing of the inputs of the ECAL portion of the first level of the trigger system in the region at $\abs{ \eta }> 2.0$ caused a specific trigger inefficiency during the 2016 and 2017 data taking, affecting photons and, to a greater extent, jets~\cite{CMS:2020cmk}.

\subsection{Photon and jet reconstruction}
\label{sec:phoJetReco}

The determination of the primary vertex from which the two photons originate has a direct impact on the diphoton invariant mass resolution. If the position along the beam axis ($z$) of the interaction producing the diphoton is known to better than approximately 10\unit{mm}, the invariant mass resolution is dominated by the photon energy resolution.  The diphoton vertex identification method is the same as in Ref.~\cite{CMS-PAS-HIG-16-040}.  A BDT is used to select a diphoton vertex from the set of all reconstructed primary vertices, incorporating as input variables the sum of the squared transverse momenta of the charged particle tracks associated with the vertex, and two variables that quantify the vector and scalar balance of \pt between the diphoton system, and the charged particle tracks associated with the vertex. Furthermore, if either photon is associated with any charged particle tracks that have been identified as resulting from conversion, the pull between the longitudinal positions of the primary vertex obtained from the conversion tracks alone and from all associated tracks is added to the BDT input variable set, as well as the number of conversions.

The photon reconstruction algorithm is the same as used in Refs.~\cite{CMS-PAS-HIG-16-040,Sirunyan:2020ycc}.  Photon candidates are reconstructed by the particle-flow algorithm~\cite{CMS-PRF-14-001}.  First, cluster ``seed" crystals are identified as local energy maxima in the ECAL above a given threshold. Second, clusters are grown from the seeds by aggregating crystals with at least one side in common with a clustered crystal and with an energy in excess of a certain threshold.  The energy of each crystal can be shared among adjacent clusters assuming a Gaussian transverse profile of the electromagnetic shower. Finally, clusters are merged into extended clusters or groups of clusters known as ``superclusters". 

The energy of a photon is computed from the sum of the energy of the clustered crystals, calibrated and corrected for changes in the ECAL response over time~\cite{Chatrchyan:2013dga}. The preshower energy is added to that of the superclusters in the region covered by this detector. To optimize the resolution, the photon energy is corrected for the containment of the electromagnetic shower in the superclusters and the energy losses from converted photons.  The correction is computed with a multivariate regression technique~\cite{CMS:EGM-14-001,Sirunyan:2020ycc} that estimates simultaneously the energy of the photon and its uncertainty.  This regression is trained on simulated photons using the ratio of the true photon energy and the sum of the energy of the clustered crystals as the criterion for optimization. The inputs are shower shapes and position variables---both sensitive to shower containment and possible unclustered energy---preshower information, and global event observables sensitive to pileup.  After the application of the regression correction, a further set of corrections derived from $\PZ\to\EE$ events where the electrons are reconstructed as photons, is applied; these align the mean of the dielectron mass spectrum in data with the expected value from simulation and also smear the resolution in simulation to match that observed in data.  Additional details concerning this procedure are contained in Ref.~\cite{CMS:2020xrn}.

Hadronic jets are clustered from particles reconstructed via the particle-flow algorithm~\cite{CMS-PRF-14-001} using the infrared- and collinear-safe anti-\kt algorithm~\cite{Cacciari:2008gp,Cacciari:2011ma}, with a distance parameter of~0.4.  Jet momentum is determined as the vectorial sum of all particle momenta in the jet, and is found from simulation to be, on average, within 5 to 10\% of the true momentum over the whole \pt spectrum and detector acceptance. Additional $\Pp\Pp$ interactions within the same or nearby bunch crossings can contribute additional tracks and calorimetric energy depositions, increasing the apparent jet momentum. To mitigate this effect, tracks identified to be originating from pileup vertices are discarded and an offset correction is applied to correct for remaining contributions~\cite{CMS:2020ebo}. Jet energy corrections are derived from simulation studies so that the average measured energy of jets becomes identical to that of particle-level jets. In situ measurements of the momentum balance in dijet, photon $+$ jet,  $\PZ + \text{jet}$, and multijet events are used to determine any residual differences between the jet energy scale in data and in simulation, and appropriate corrections are made~\cite{Khachatryan:2016kdb}. Additional selection criteria are applied to each jet to remove jets potentially dominated by instrumental effects or reconstruction failures~\cite{CMS:2020ebo}. 

\subsection{Event selection and classification}
\label{sec:selection}

Photon candidates are subject to
a preselection
that imposes requirements
on
\pt, H/E, and shower shape, and that
uses an
electron veto
to reject photon candidates geometrically matched to at least two hits in the pixel detector.  
The electron veto is reinforced by an additional requirement that no photon candidate be also reconstructed as an electron candidate, which can happen if the candidate is associated with a track containing a single hit in the first layer of the pixel detector~\cite{Sirunyan:2020ycc}.

This preselection is designed to be slightly more stringent than the trigger requirements.
A photon identification BDT,
trained with the \textsc{tmva} package~\cite{tmva}, combining lateral shower shape and isolation variables~\cite{Sirunyan:2020ycc},  the median energy density, the pseudorapidity, and the raw energy is used to separate prompt photons from nonprompt photons
resulting mainly from neutral-meson decays~\cite{HIG-13-001,CMS-PAS-HIG-16-040}.  Each photon candidate must satisfy
the preselection
requirements as well as a
requirement on the minimum value of the photon identification BDT output score.  The efficiencies of the minimum photon identification BDT output score requirement and preselection criteria (except for the electron
veto requirements) 
are measured with a tag-and-probe
technique applied to $\PZ\to\EE$ events. The
fractions of photons that satisfy the two electron-veto requirements are measured with
$\PZ\to\MM\gamma$ events, in which the photon is produced by final-state
radiation
providing a sample of prompt photons with purity higher than 99\%. The ratios of the efficiencies in data and simulation
are used to correct the signal efficiency in simulated signal samples and the associated statistical and systematic uncertainties are propagated to the expected signal yields.

The analysis uses all events
that contain a diphoton pair where each of the photons in the pair
satisfy a requirement
on 
the ratio of its
\pt value to the invariant mass of the diphoton system, \mgg, that must lie within the range $65<\mgg<120\GeV$.  Specifically, the requirements are $\ptga/\mgg>0.47$ and  $\ptgb/\mgg>0.28$.
Here, $\gamma 1$ ($\gamma 2$) refers to the photon candidate in the pair with the higher (lower)
\pt value.
The use of \pt thresholds scaled by \mgg is intended to prevent
a distortion of the low end of the
diphoton mass spectrum that results if fixed thresholds are used~\cite{HIG-13-001,CMS-PAS-HIG-16-040};
in particular, the minimum \pt values in the above fractions, 30.6 and 18.2\GeV, are chosen to be slightly higher than those of the HLT paths, \ie, 30 and 18\GeV, to further guard against
distortion of the spectrum.  To further discriminate against surviving electron pairs from DY events, we require that the natural logarithm of the sum of the squares of transverse momenta of all tracks associated with the chosen diphoton vertex, $\ln(\Sigma\pt^{2}/\GeV^2)$, satisfy the requirement $\ln(\Sigma\pt^{2}/\GeV^2)<0.016\pt^{\gamma\gamma}/\GeV + 6.0$, where $\pt^{\gamma\gamma}$ is the transverse momentum of the selected diphoton candidate in each event. This requirement, optimized on surviving simulated DY and signal events, acts against tracks from DY electrons, the squared \pt of which are included in this sum, while preserving the efficiency of signal events, in particular those where the diphoton system is boosted.  The efficiencies of this requirement in data and simulation are measured with $\PZ\to\EE$ events, where the tracks matched to each of the two photon candidates have been removed. As with the preselection requirements, the data/simulation efficiency ratio is used to correct the signal sample efficiency, and the associated uncertainties are likewise propagated to the expected signal yields. 

A multivariate event classifier~\cite{HIG-13-001,CMS-PAS-HIG-16-040}, the diphoton BDT, 
is
used to discriminate between diphoton events from Higgs boson decays
and those from the diphoton continuum, to further reduce background from events containing jets misidentified as isolated photons, and to assign higher scores to events with good diphoton mass resolution.
Trained with the \textsc{tmva} package, it incorporates the kinematic properties of the diphoton system (excluding
\mgg), a per-event estimate of the diphoton
mass resolution, and the photon identification BDT output values.

To target events produced via the VBF process, which feature two jets in the final state separated by a large rapidity gap, an additional multivariate discriminant~\cite{HIG-13-001,CMS-PAS-HIG-16-040} is trained with the \textsc{Scikit-learn}  package~\cite{Pedregosa:2011ork}, for the analysis of the 2017 and 2018 data, to identify the distinctive kinematics of these jets, considering the $\Pg\Pg\PH$ + jets production process as background. The output score of this 
discriminant is given as input to a further multivariate classifier~\cite{HIG-13-001,CMS-PAS-HIG-16-040} trained with \textsc{Scikit-learn}, the ``combined" BDT, along with the score from the diphoton BDT, and the ratio of the \pt of the diphoton system to its invariant mass, $\pt^{\gamma\gamma}/\mgg$. 

As the first step of the classification process, events from the 2017 and 2018 data sets having two jets identified as not being compatible with pileup~\cite{CMS-PAS-HIG-16-040}, satisfying the requirements $\pt^{j_{1}}(\pt^{j_{2}})>30\,(20)\GeV$, $\abs{\eta_{j_{1}(j_{2})}}<4.7$,
where $j_1$ ($j_2$) refers to the jet with the highest (next-highest)
\pt value, $m_{j_{1}j_{2}}>100\GeV$, and having a minimum  
combined BDT classifier score value, are assigned to a ``VBF class".
The remaining events, as well as all those from the 2016 data set, are separated into classes based on the diphoton BDT score, with a minimum score below which they are rejected.  The number of these classes and their boundaries
are determined so as to maximize the expected signal significance for $\mH=90\GeV$, the midpoint of our search range, while providing a sufficient number of simulated events from \PZ boson dielectron decays for background modeling; different boundaries are used for the analysis of the 2016 data than for that of 2017 and 2018.  Three such classes are used; they are referred to as 0, 1, 2, where class 0 contains the events with greatest expected sensitivity.
The fraction of events containing more than one diphoton candidate is of order $10^{-4}$. In these cases, the candidate assigned to the highest sensitivity class is selected; should this class still contain multiple diphoton candidates, the candidate with the highest value of $\ptga+\ptgb$ is then selected.
\section{Signal parametrization}
\label{sec:signal}
In order to perform a statistical interpretation of the
data, it is necessary to have a description of the signal
that
includes the overall product of the efficiency and acceptance,
as well as the shape of the diphoton mass distribution in each of the
event classes. The simulated SM-like Higgs boson events are used
to
construct a parametrized signal model
that is defined continuously for any value of
the Higgs boson mass between 70 and 110\GeV.  As mentioned in Section~\ref{sec:phoJetReco}, the photon energy resolution predicted by the simulation is modified by a Gaussian smearing determined from the comparison between the $\PZ\to\EE$ line-shape in data and simulation, where the electron energies have been corrected with factors developed for photons, using the same
procedure as that described in Refs.~\cite{HIG-13-001,CMS-PAS-HIG-16-040}.  The amount of smearing is extracted differentially in  bins of $\abs{\eta}$ and
the $\RNINE$ shower shape variable~\cite{CMS:EGM-14-001},
defined as the energy sum of $3{\times}3$ crystals centered on the most energetic crystal in the ECAL cluster divided by the energy of the cluster.  The trigger and preselection efficiency corrections described in Sections~\ref{sec:triggerSim} and~\ref{sec:selection}, respectively,
are also applied to the simulated signal events.

Since the shape of the \mgg distribution changes considerably depending on whether the vertex associated with the candidate diphoton is correctly identified, separate fits are made to the distributions for the correct and incorrect primary vertex selections when constructing the signal model.  Events are considered to have the correct primary vertex if the $z$ coordinate of the vertex associated with the candidate diphoton is within 1\unit{cm} in distance along the beam axis of the true vertex.  For these events the signal shape
is dominated by the 
ECAL response and reconstruction, and
is modeled empirically with a sum of between two and four Gaussian functions depending on the data-taking period (2016, 2017, and 2018) and event class. The signal shape for events with an incorrect primary vertex selection is broadened significantly by the difference between the $z$ coordinate position of the selected primary vertex and that of the true Higgs boson production
vertex.  The signal shape for these events is modeled with a sum of up to three Gaussian functions, depending on the data-taking period and
event class. In
both the correct and incorrect vertex assignment cases, the number of Gaussian functions is determined using an F-test~\cite{ftest}, and the means, widths, and relative fractions of the Gaussian functions are extracted from the fits.
For each production process, event class, and primary vertex identification correctness scenario, a simultaneous fit of signal samples~\cite{CMS-PAS-HIG-16-040} at \mH values in the range from 70 to 110\GeV is performed to obtain parametric variations with \mH, described by polynomials, of the Gaussian function parameters.  The final fit function for each event class in each production process and each year is obtained by summing the individual functions for the two vertex scenarios. The relative size of the shapes associated with the two vertex scenarios, in each final function, is determined by the vertex selection efficiency. 

\begin{figure*}[htb!]
 \centering
   \includegraphics[width=0.49\textwidth]{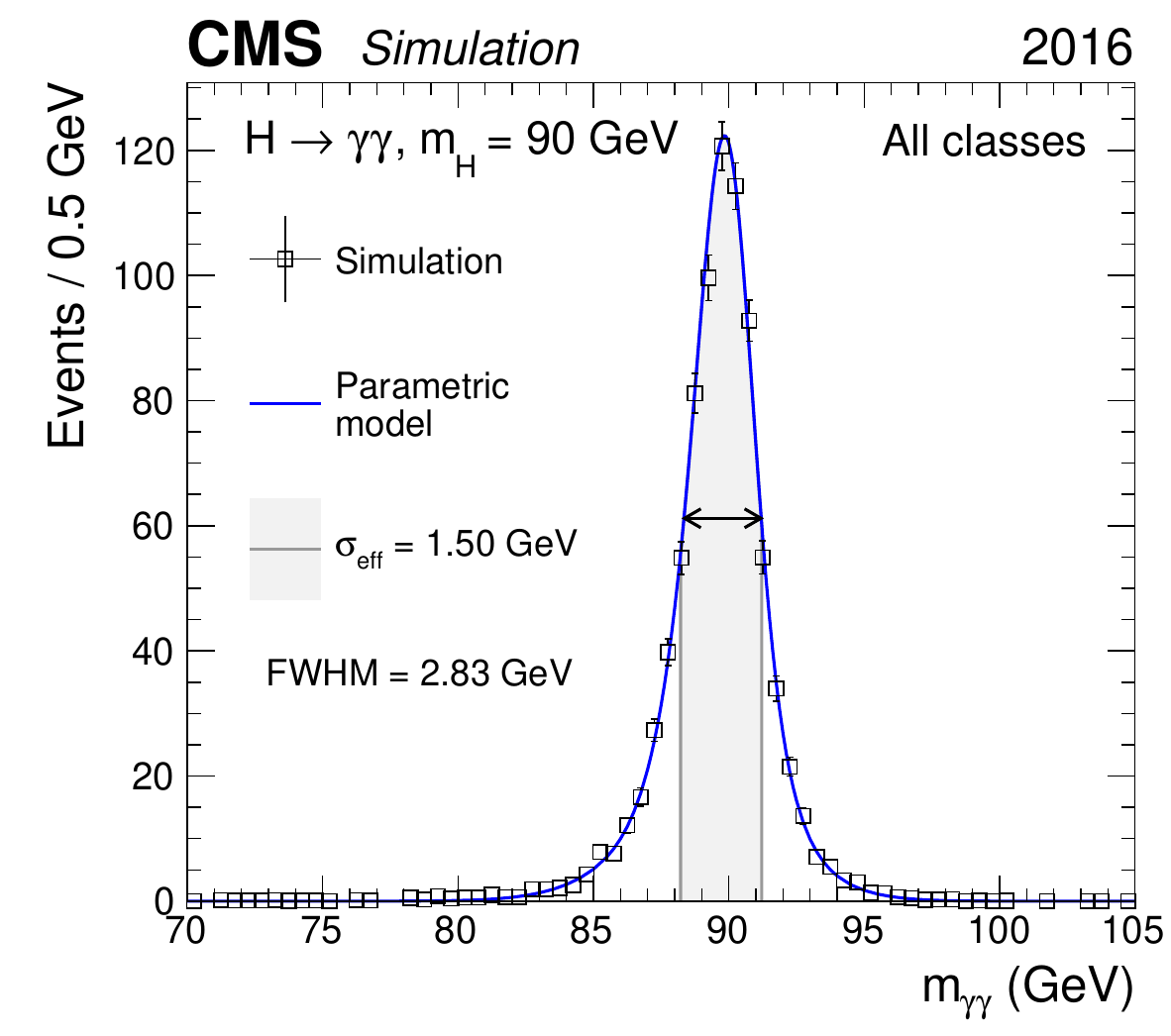}
   \includegraphics[width=0.49\textwidth]{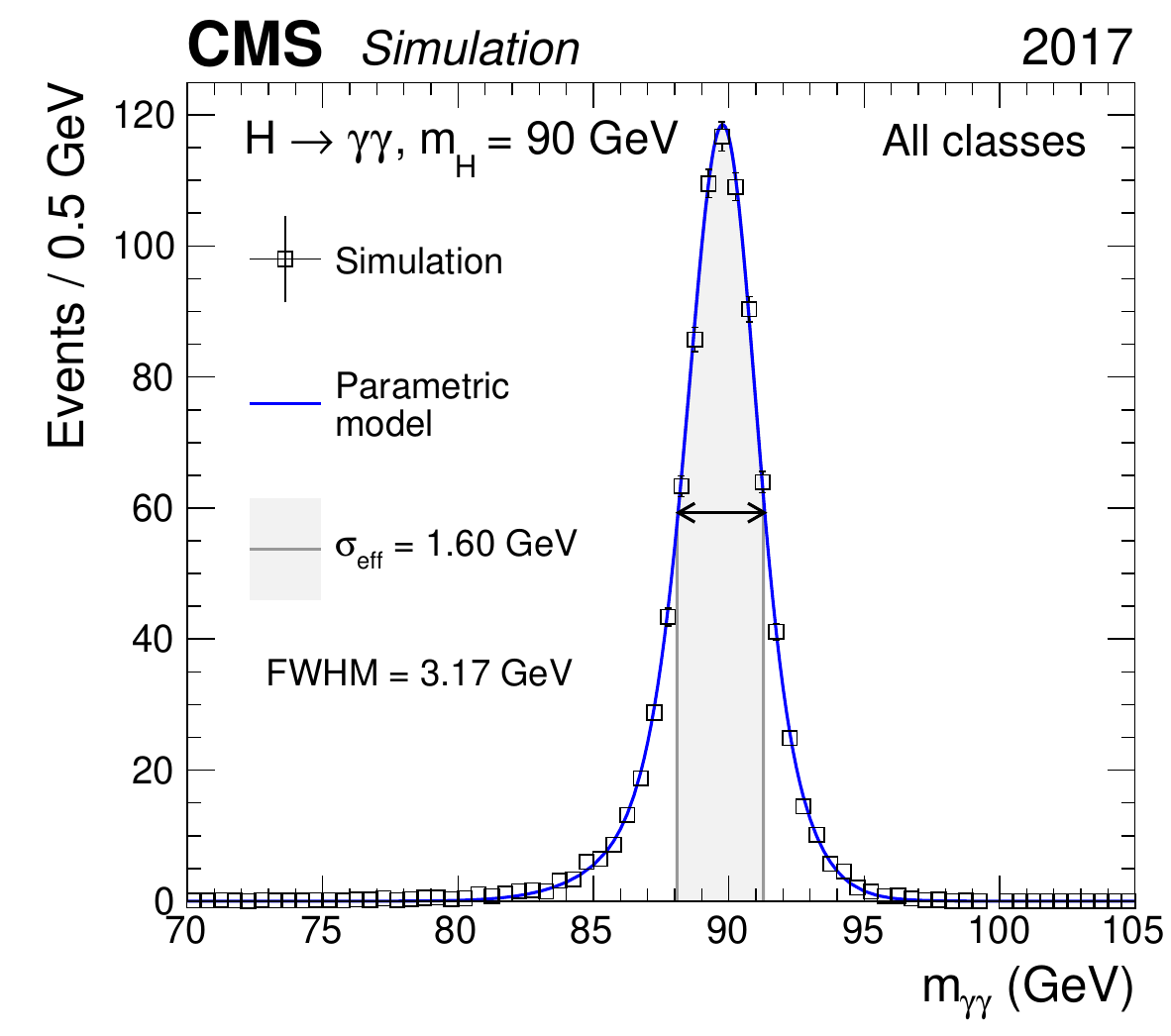} \\
   \includegraphics[width=0.49\textwidth]{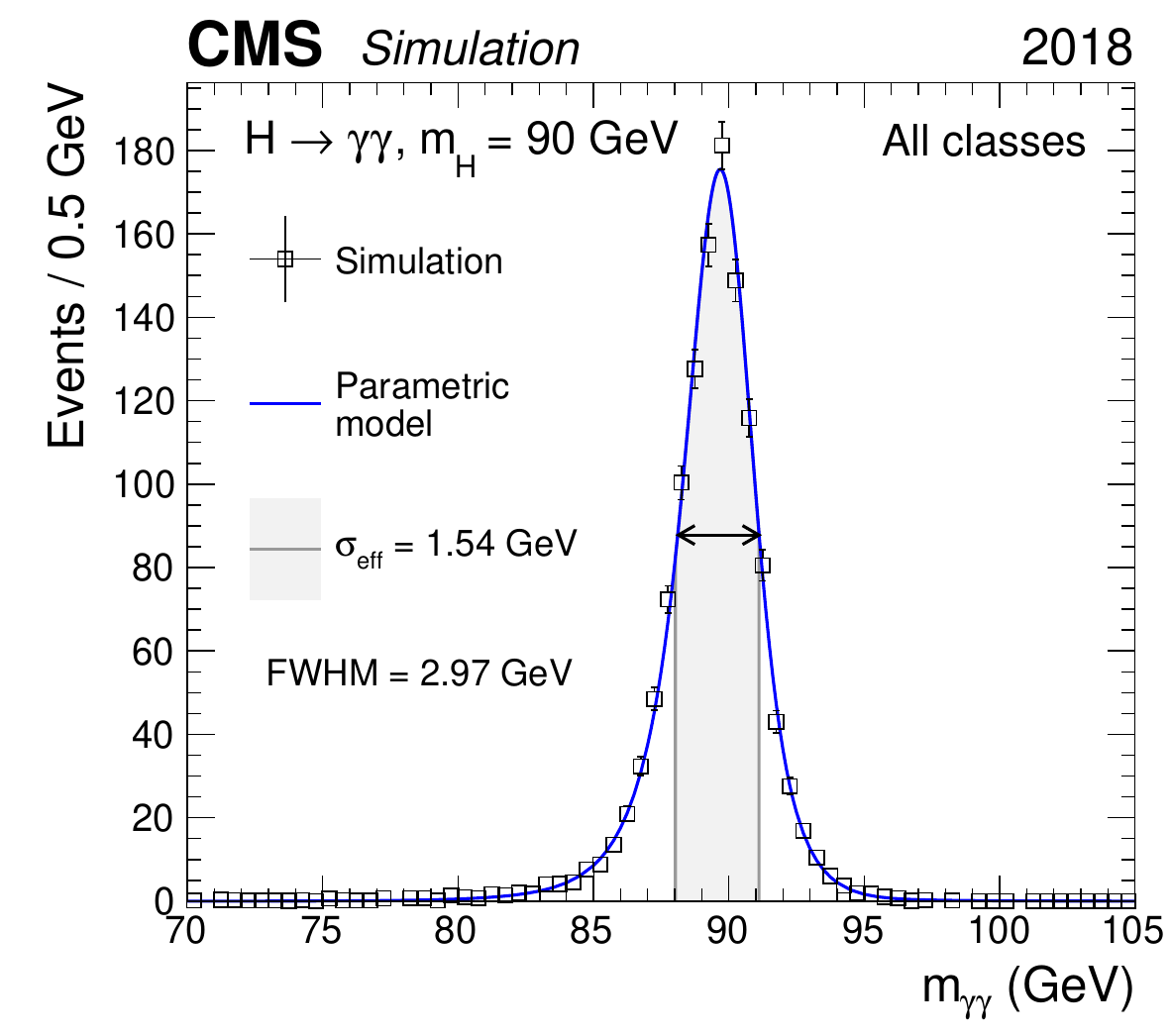}
\caption{Full parametrized signal shape, integrated over all event classes, in simulated signal events with $\mH=90\GeV$ for 2016 (upper left), 2017 (upper right), and 2018 (lower). The open points are the weighted MC events and the blue lines the corresponding parametric models. Also shown are the  $\sigma_{\text{eff}}$ values and the shaded region limited by $\pm\sigma_{\text{eff}}$, along with the FWHM values, indicated by the position of the arrows on each distribution. \label{sigPlotsComb}}
\end{figure*}

The final parametrized shapes
for the combination of all production mechanisms, for all event classes, weighted by their SM cross sections
are shown in Fig.~\ref{sigPlotsComb} for a Higgs boson mass of
90\GeV for the 2016, 2017, and 2018 data.  Also shown are the full width at half maximum (FWHM) value and
the value of the effective
standard deviation ($\sigma_{\text{eff}}$), which is defined as half the width of the narrowest interval containing 68.3\% of the
invariant mass distribution.
The product of efficiency and acceptance of the signal model ranges from 17.5 
(16.0 and 16.9)\% 
for $\mH=70\GeV$ to 20.8 
(19.9 and 20.8)\% 
for $\mH=110\GeV$ in the case of the 2016 (2017 and 2018) data.

\section{Background estimation}
In this analysis, as in Refs.~\cite{Chatrchyan:2012xdj,HIG-13-001,CMS-PAS-HIG-16-040},
the background is modeled by fitting analytic functions to the observed diphoton mass distributions in each of the event classes.
The fits are performed over the range $65 < \mgg < 120\GeV$.  The model is
determined from data with the discrete profiling method~\cite{DiscreteProfilingMethod}, which treats the choice of the background function as a discrete parameter in the likelihood fit to the data and estimates the systematic uncertainty associated with the choice of a particular function.

Since the search mass range of this analysis includes the
\PZ boson peak region, a significant potential background source is DY dielectron production that, through misidentification, could appear as two isolated photons.  An explicit component intended to describe the background from the DY process, in which the two apparently isolated photons 
survive all the selection requirements stated in Section~\ref{sec:selection}, is added to the smoothly falling function~\cite{Chatrchyan:2012xdj, HIG-13-001, CMS-PAS-HIG-16-040}
used to model the background.  An exception is made for the events of the VBF class, where the requirement on the minimum value of the combined BDT classifier is not imposed in order to obtain a sufficient number of events for the modeling.  This additional component, referred to as 
``doubly-misidentified" events, is modeled
with
a double-sided Crystal Ball (DCB) function, which is a  modification of the Crystal Ball function~\cite{CrystalBall,DoubleCrystalBallRef} with an exponential tail on
both sides, summed with an additional exponential function (DCB + exponential).
The DCB function
is characterized by seven parameters, including
the number of events
for normalization, and six shape parameters:  the Gaussian mean and standard deviation, $\alpha_{\text{L}}$, $n_{\text{L}}$, $\alpha_{\text{R}}$, and $n_{\text{R}}$, where
$\alpha_{\text{L,R}}$ and $n_{\text{L,R}}$ refer, respectively, to the slope and normalization of the left-hand (L) and right-hand (R) exponential tails.  The additional exponential function contributes two supplementary parameters:  $p$, the coefficient of the exponential argument and $g$, the fraction that the exponential function represents in the total DY process model.  
The values of these two parameters as well as those describing  
the DCB
shape are
determined by fitting the diphoton invariant mass distribution in
a sample of
simulated doubly-misidentified DY events for each event class.
Because of the
small size of the simulated event
sample, we fix two of the six DCB
shape parameters, $\alpha_{\text{L}}$ and $\alpha_{\text{R}}$,
to make the fit more stable. The fixed values are different in each event class
and are obtained from fits of events where one photon candidate survives all selection requirements including the electron vetoes, and the other survives all selection requirements
but fails the electron vetoes (``singly-misidentified" events).  In each class, the value of the
mean, which coincides with the peak position, lies somewhat below the nominal \PZ boson mass value. This bias is due to
the fact that the electrons surviving the photon selection requirements (in particular the electron veto) have in general been poorly reconstructed, for example having undergone wide-angle bremsstrahlung of high-energy photons;
furthermore, the electron energies have been corrected
with factors developed for photons.

For the application of the discrete profiling method, members of several families of analytic functions, including
exponential,
power law,
polynomials in the Bernstein basis,
and Laurent
series, are considered, each summed with the DCB + exponential function.  The
maximum order term in each series
is determined using an F-test~\cite{ftest}, and the minimum order, using
a goodness-of-fit test.  When fitting these functions to the \mgg distribution in the data, the value of twice the negative logarithm of the likelihood (2NLL) is minimized.  A penalty is added to the 2NLL value to take into account the number of floating parameters, including
the fraction $f$ of background events attributed to the component arising from the doubly-misidentified events (DCB + exponential fraction), in each candidate function.

The normalization of the DY background is
determined from the fit.
The shape parameters are fixed to the constant values
that are obtained by fitting the
doubly-misidentified DY events, as described above.   In particular, the value of the Gaussian standard deviation in each event class for these doubly-misidentified events is greater than the corresponding value of $\sigma_{\text{eff}}$ in the signal model by a factor of up to 2.

Table~\ref{tab:bckdOnlyModelsPlusDCBExpFractions} shows, for each event class in each data-taking year, the functional family and order of the function chosen by the 2NLL minimization as being the best fit when summed with the DCB + exponential function, in the case of a background-only fit.  Also shown are the DCB + exponential fractions for these chosen models in the subset of the diphoton mass range extending from 85 to 95\GeV, the mass range where the expected dielectron background from DY processes is most significant.  These fractions range from 0.5 to 4.8\% across the different years and event classes.

\begin{table*}[!tb]
\centering
\topcaption{Families and orders of functions chosen as best fit when summed with the DCB + exponential function, by year and by event class, in the case of background-only fits.  The DCB + exponential fractions for these models in the range $85<\mgg<95\GeV$ are also shown. 
}
\cmsTable{
\begin{tabular}{lrrrrr}
\hline
Event class & & 0 & 1 & 2 & VBF \\
\hline
2016 & Family/Order &  Power Law 1  &  Bernstein 4 &  Exponential 3  &  \NA  \\
 & DCB + Exp. Fraction (\%) &   3.0  &  3.1  &  3.3 & \NA  \\
[\cmsTabSkip]
2017 & Family/Order  & Bernstein 3  &  Exponential 3  &  Bernstein 4  &  Bernstein 3   \\
 & DCB + Exp. Fraction (\%)  &  2.7  &  1.4  &  1.9  &  2.6 \\
[\cmsTabSkip]
2018 & Family/Order   &  Laurent 1  &  Bernstein 4  &  Exponential 3  &  Bernstein 2  \\
 & DCB + Exp. Fraction (\%)   &   0.5  &  4.1  &  4.8  &  0.8\\
\hline
\end{tabular}
}

\label{tab:bckdOnlyModelsPlusDCBExpFractions}
\end{table*}

\begin{figure*}[!thb]\centering
   \includegraphics[width=0.49\textwidth]{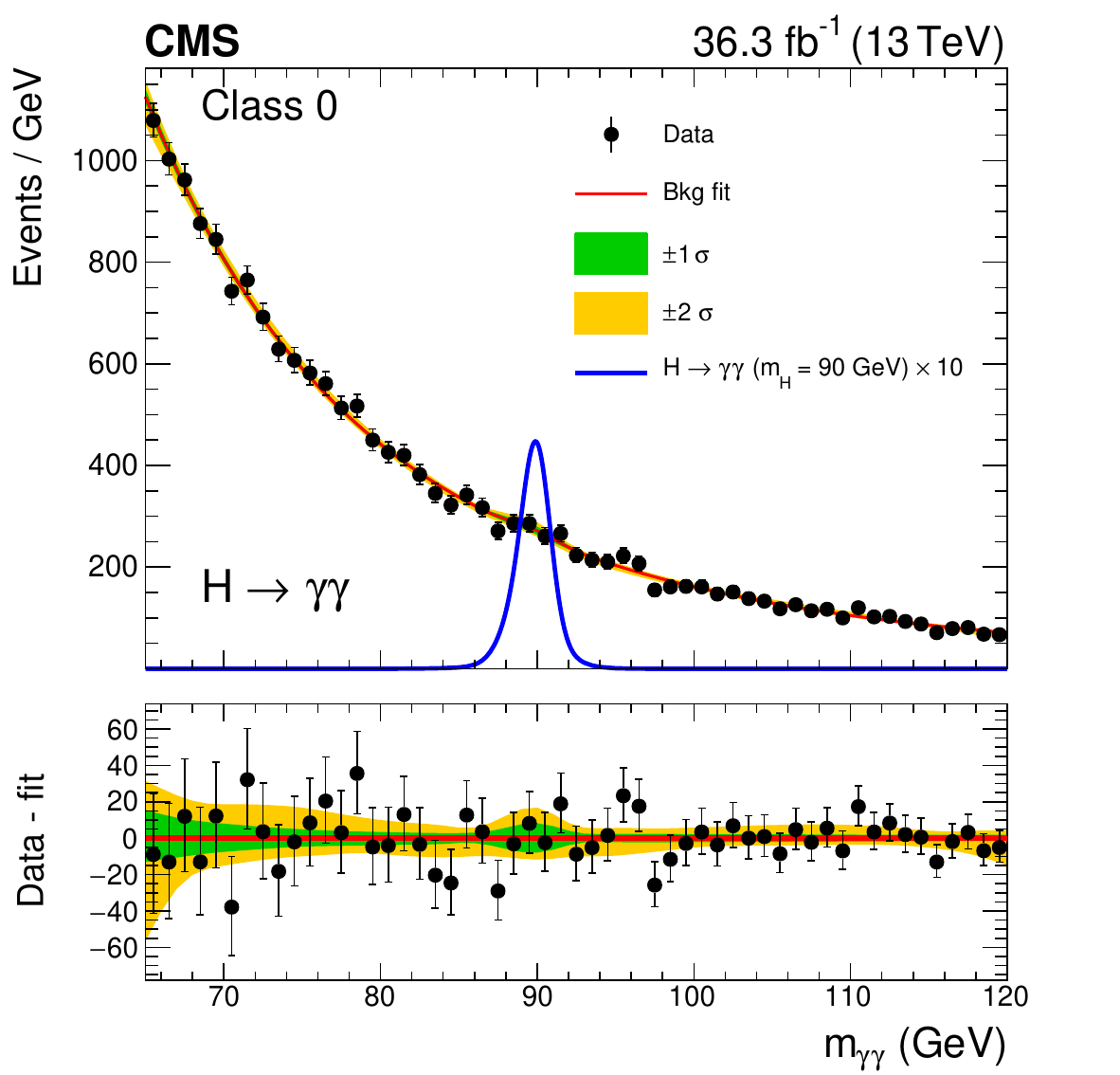}
   \includegraphics[width=0.49\textwidth]{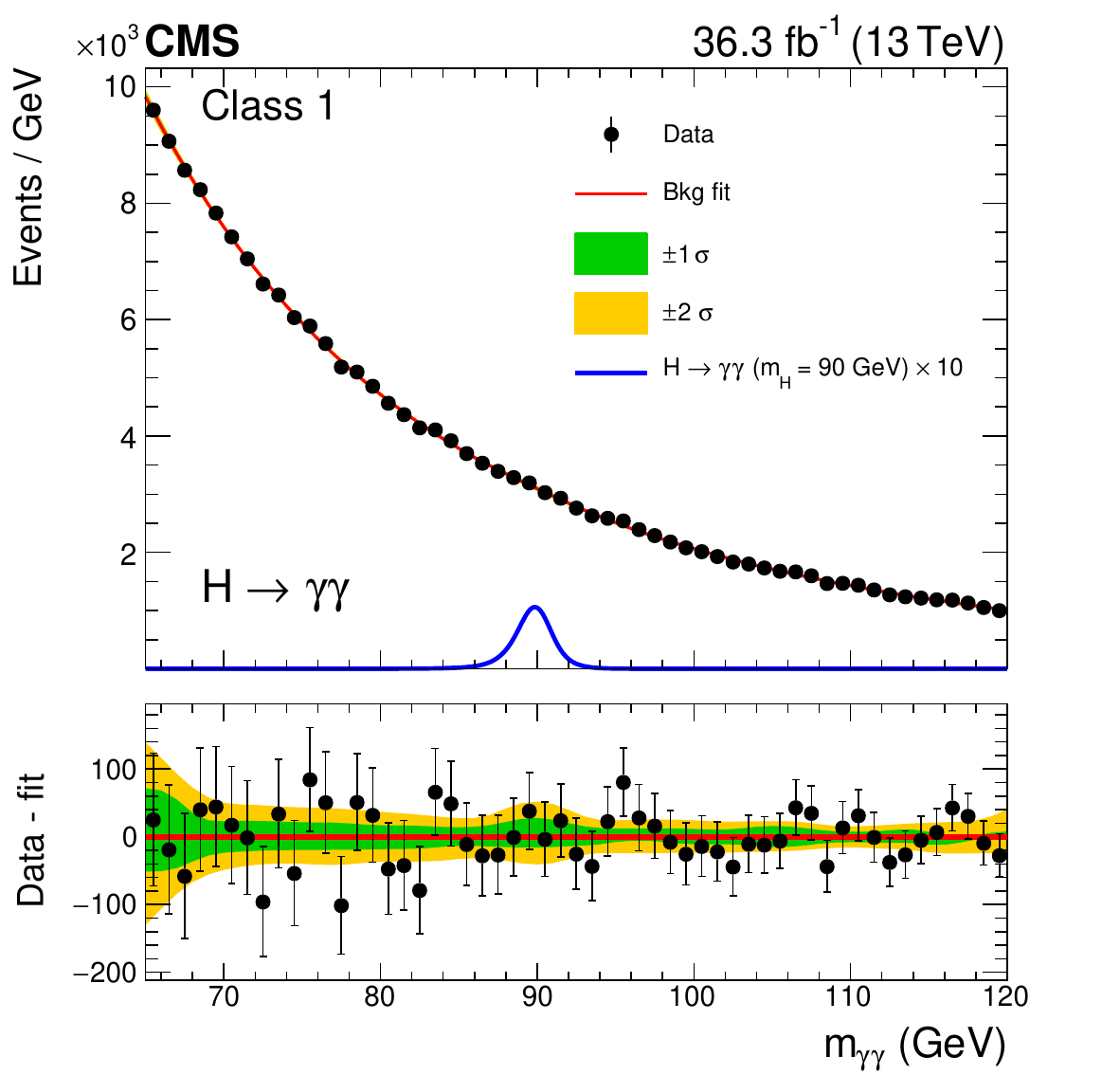} \\
   \includegraphics[width=0.49\textwidth]{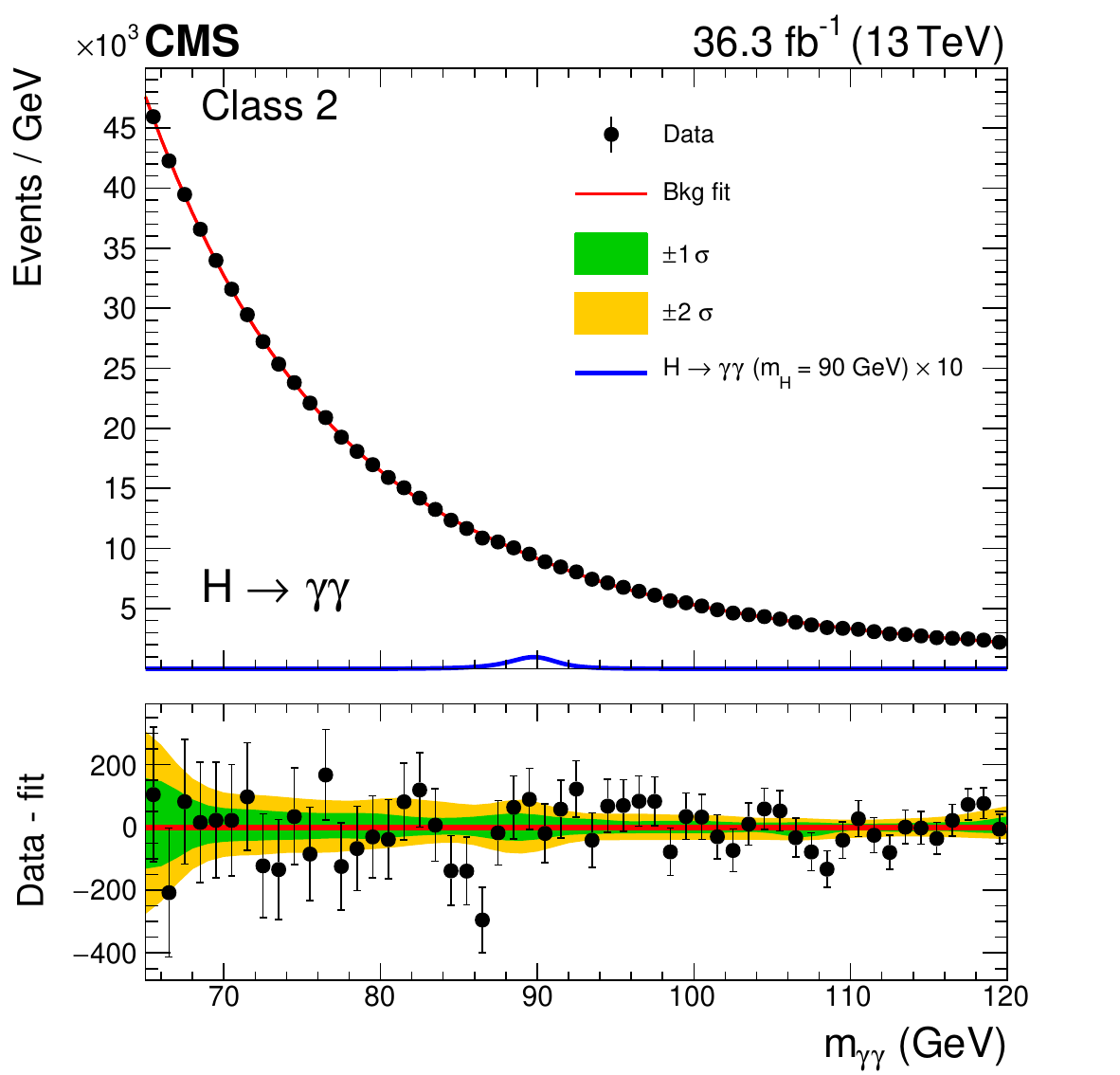}
    \caption{Background model fits using the chosen background model parametrization to the 2016 data in the three event classes.  The corresponding signal model for each class and $\mH=90\GeV$, multiplied by 10, is also shown. The one- and two-$\sigma$ bands reflect the uncertainty in the background model normalization associated with the statistical uncertainties of the fits, and are shown for illustration purposes only. The difference between the data and the background model is shown in the lower panels.}
     \label{bkgModelFinal2016}
   \end{figure*}

\begin{figure*}[!thb]\centering
   \includegraphics[width=0.49\textwidth]{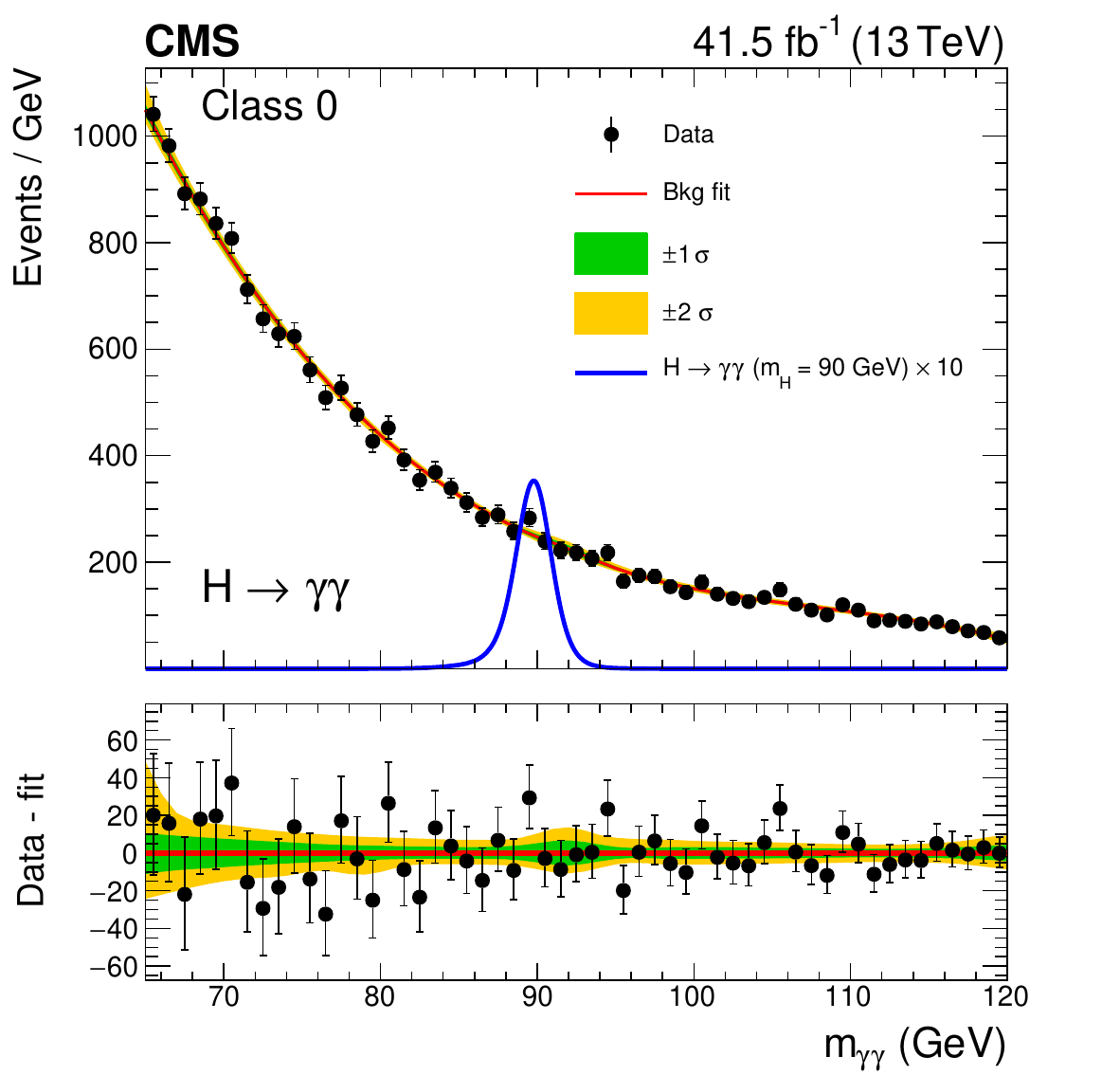}
   \includegraphics[width=0.49\textwidth]{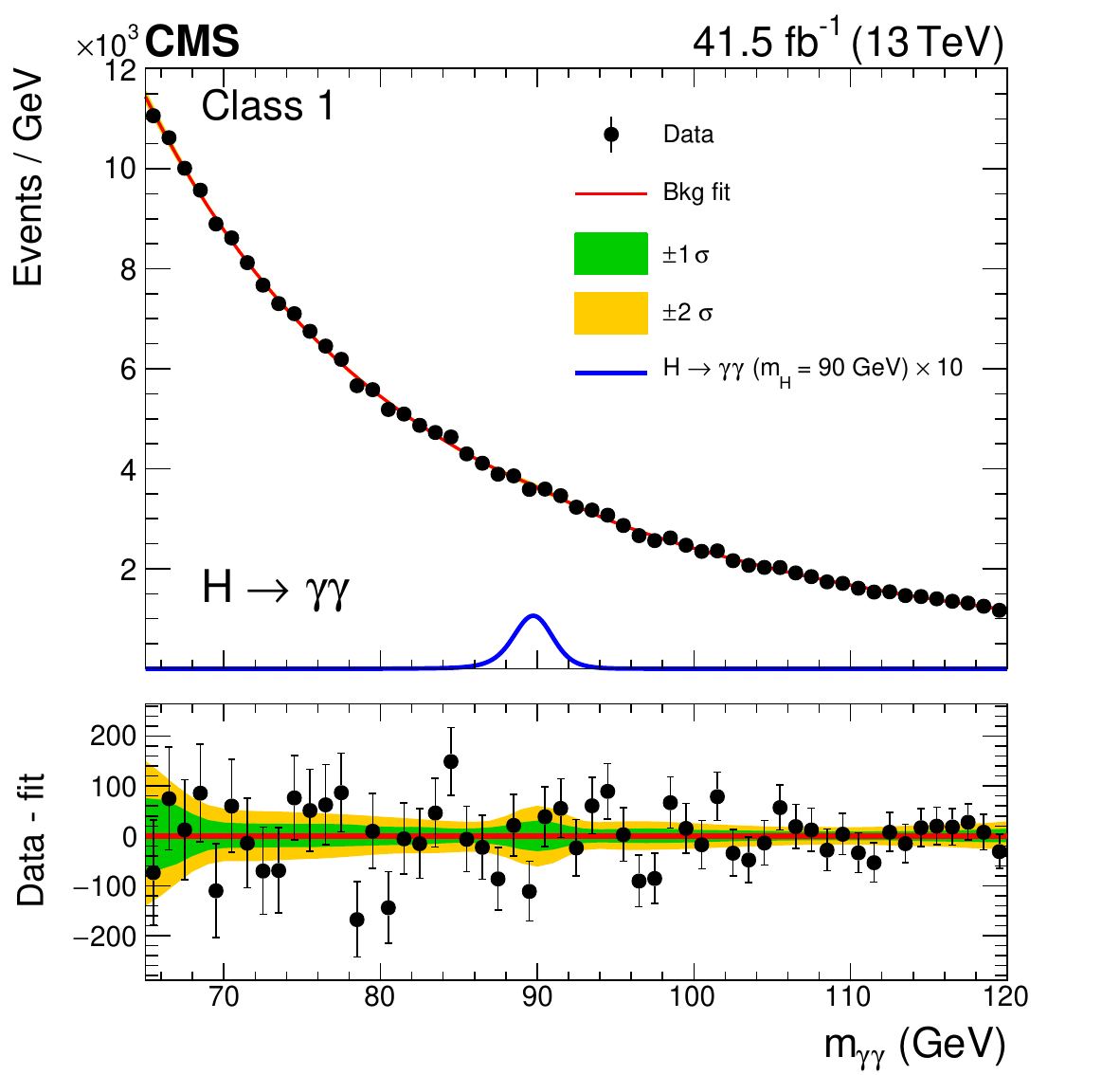} \\
   \includegraphics[width=0.49\textwidth]{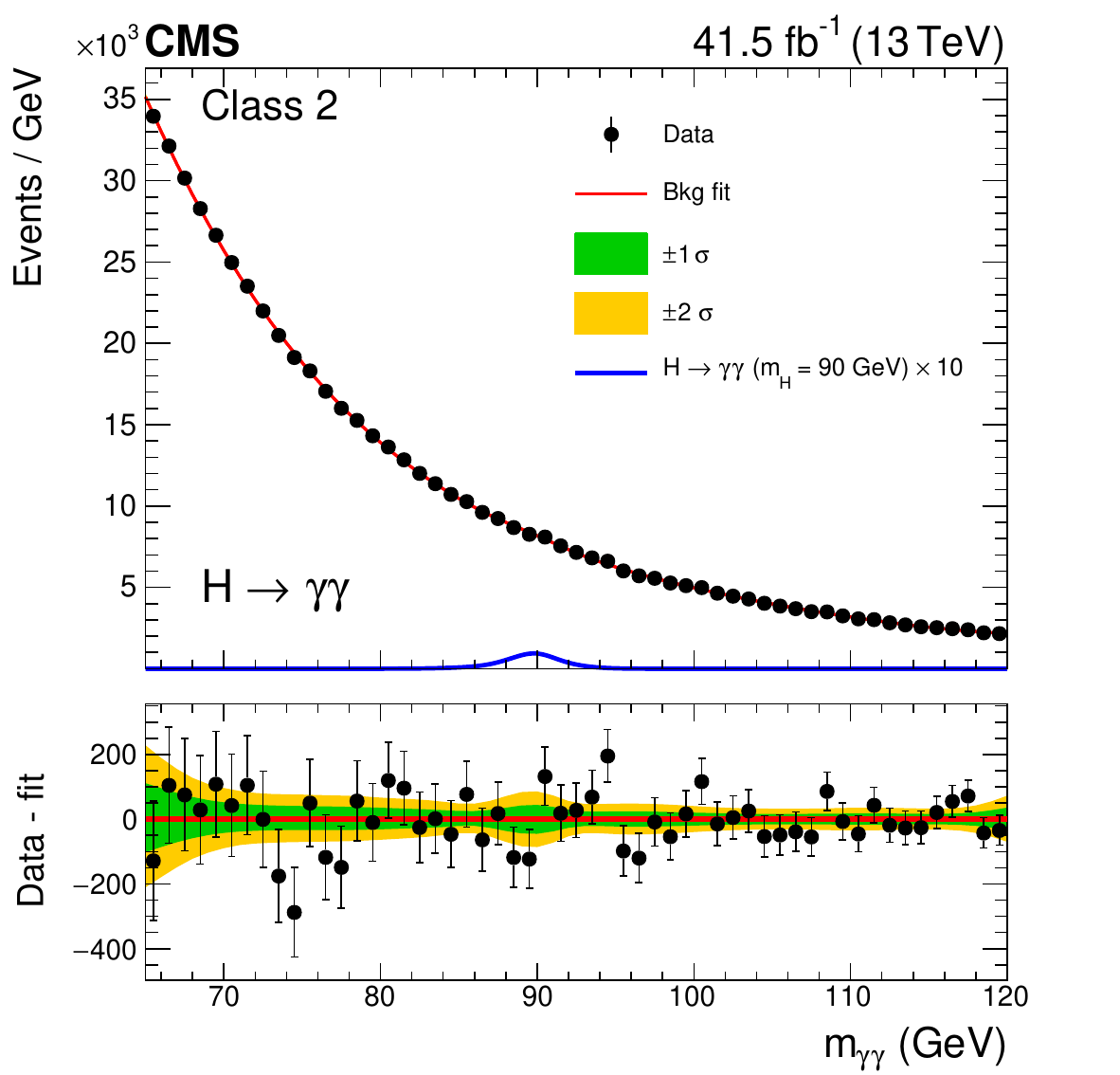}
   \includegraphics[width=0.49\textwidth]{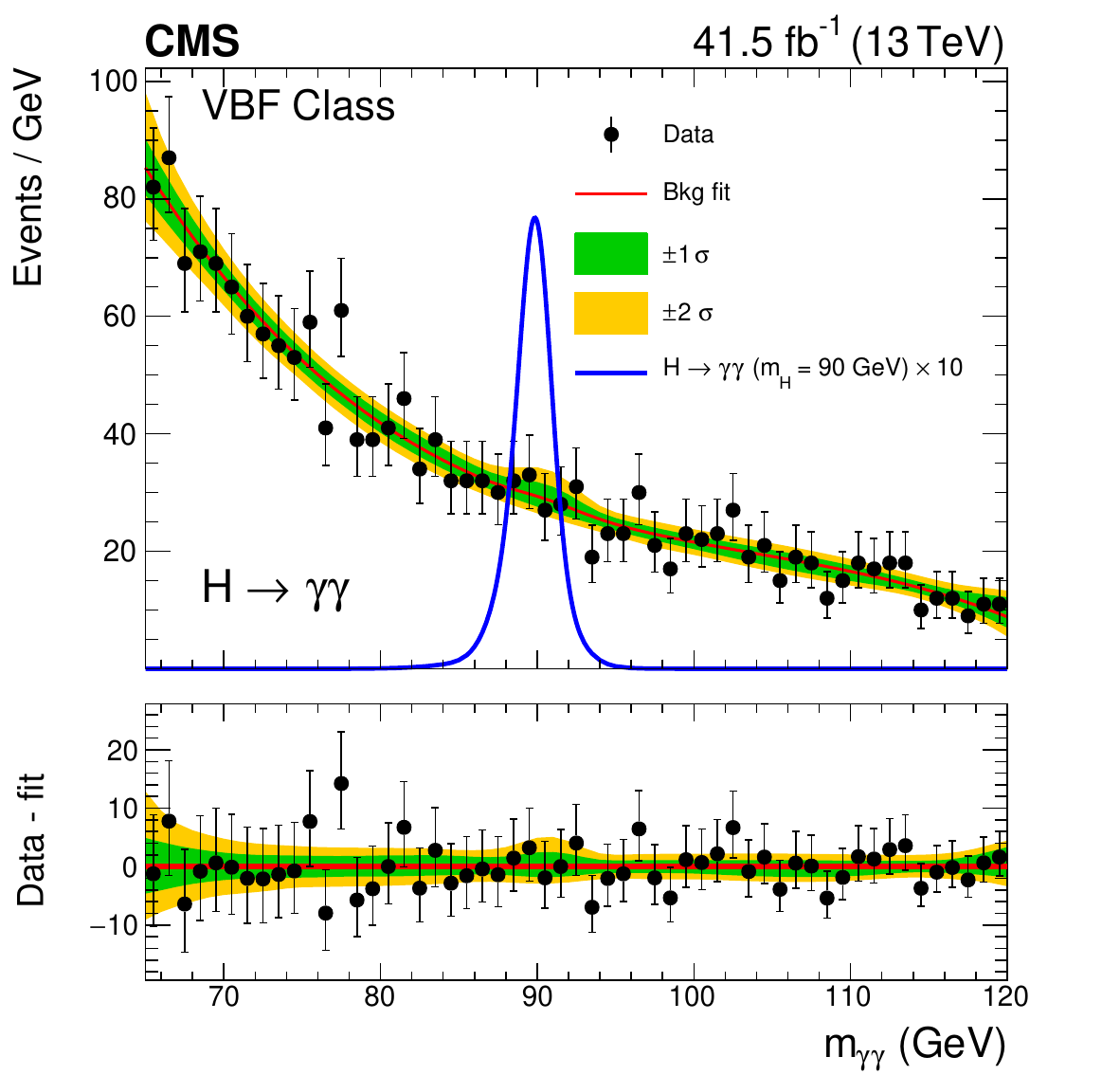}
   \caption{Background model fits using the chosen background model parametrization to the 2017 data in the four event classes.  The corresponding signal model for each class and $\mH=90\GeV$, multiplied by 10, is also shown. The one- and two-$\sigma$ bands reflect the uncertainty in the background model normalization associated with the statistical uncertainties of the fits, and are shown for illustration purposes only. The difference between the data and the background model is shown in the lower panels.}
     \label{bkgModelFinal}
   \end{figure*}

\begin{figure*}[!htb]\centering
 \includegraphics[width=0.49\textwidth]{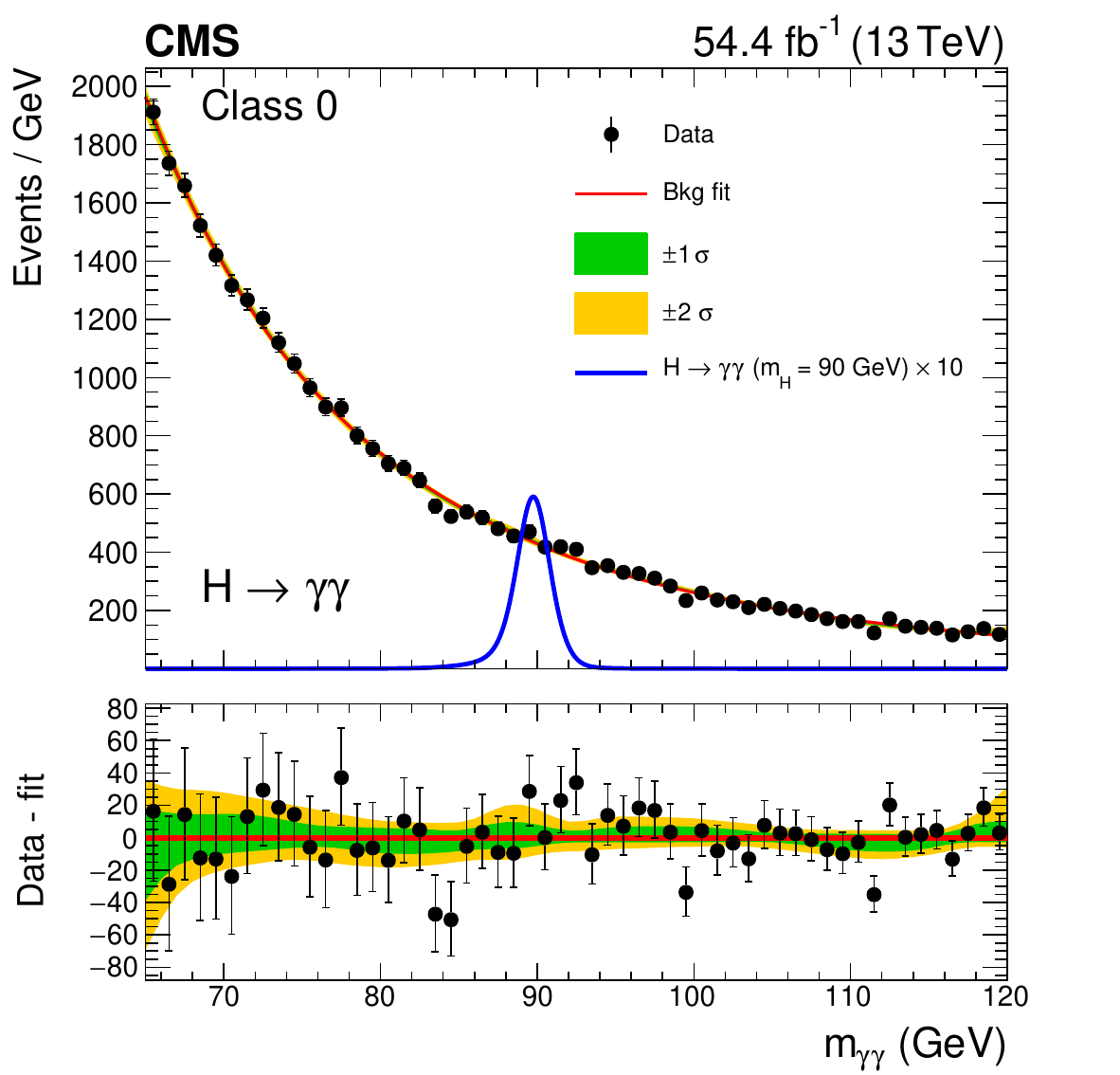}
 \includegraphics[width=0.49\textwidth]{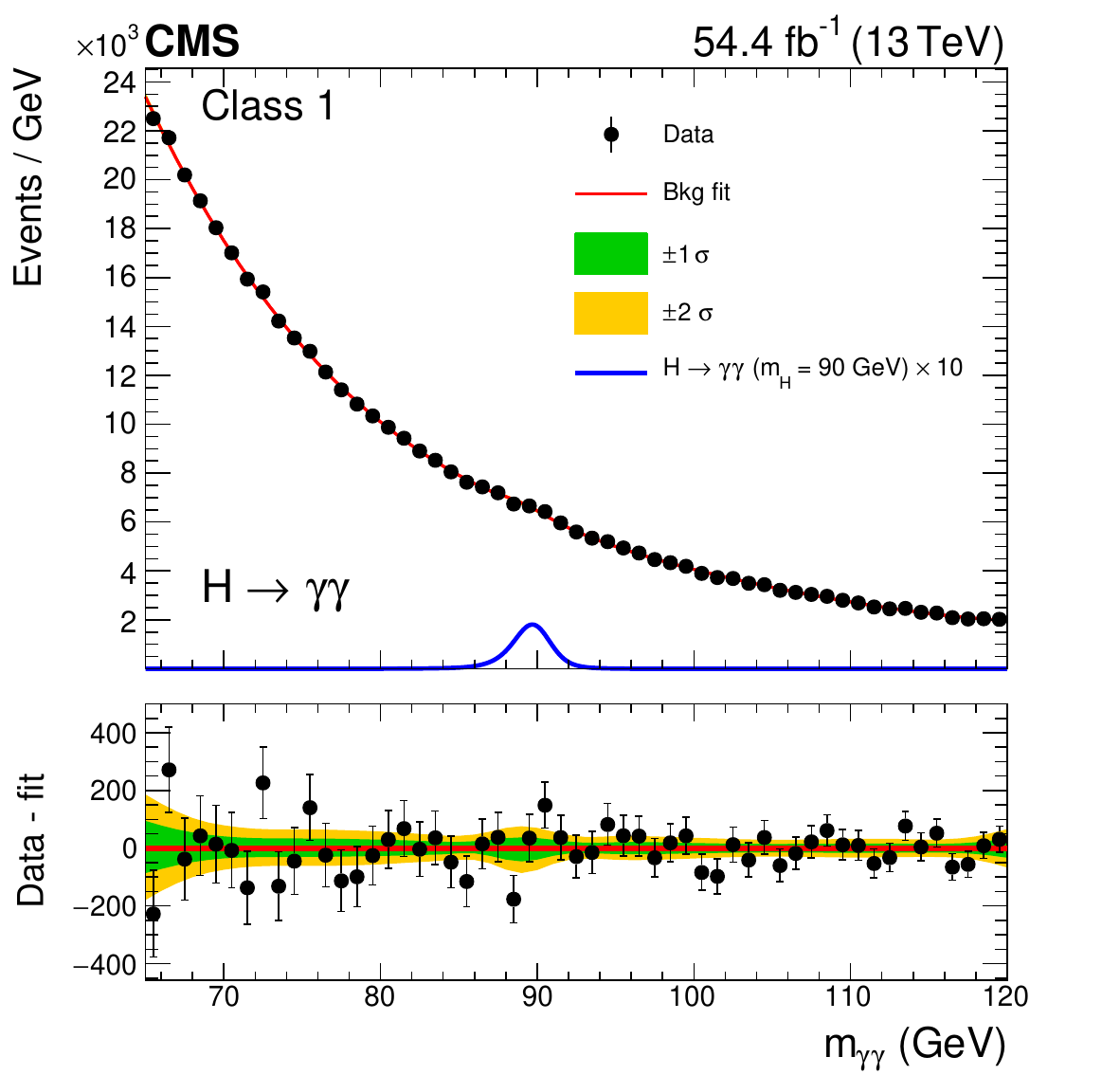} \\
 \includegraphics[width=0.49\textwidth]{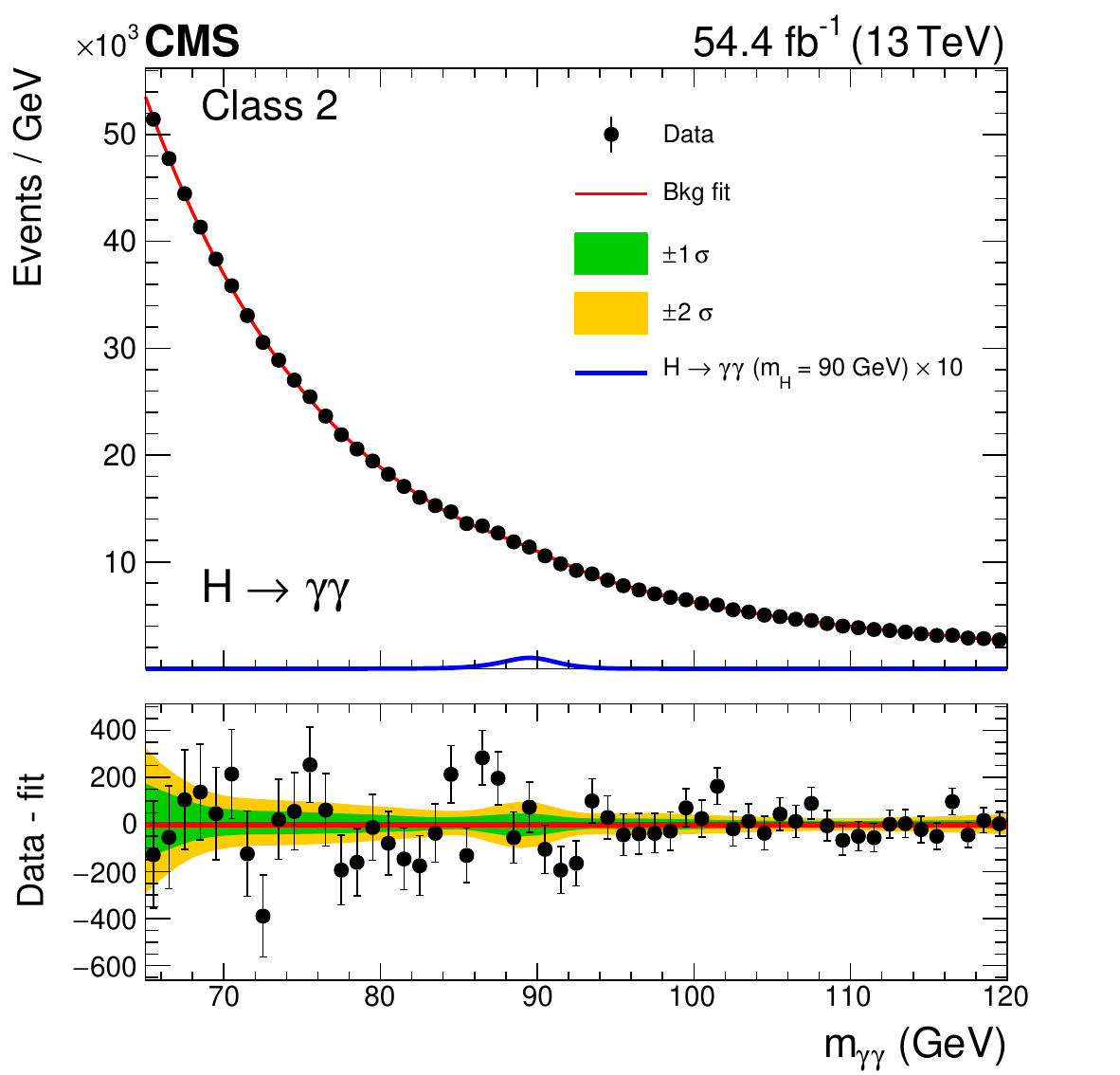}
 \includegraphics[width=0.49\textwidth]{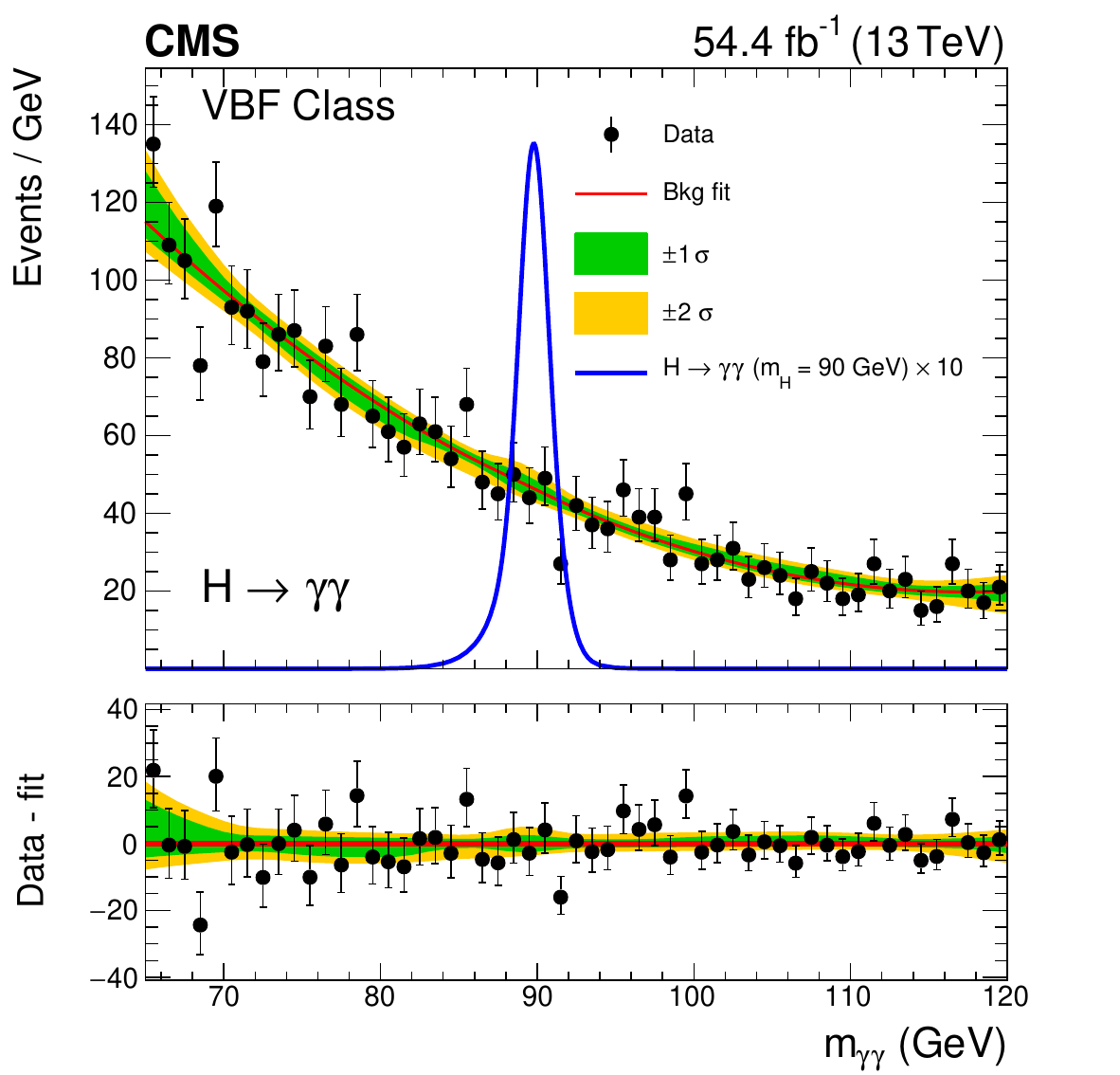}
 \caption{Background model fits using the chosen background model parametrization to the 2018 data in the four event classes.  The corresponding signal model for each class and $\mH=90\GeV$, multiplied by 10, is also shown. The one- and two-$\sigma$ bands reflect the uncertainty in the background model normalization associated with the statistical uncertainties of the fits, and are shown for illustration purposes only. The difference between the data and the background model is shown in the lower panels.}
 \label{fig:statAnalysisBkgValidationPlotsUntagged}
\end{figure*}

Binned likelihood fits of the chosen background models to the observed \mgg distributions are performed, using 1\GeV bins and 
assuming no signal, and are shown for all the event classes in Figs.~\ref{bkgModelFinal2016},~\ref{bkgModelFinal}, 
and~\ref{fig:statAnalysisBkgValidationPlotsUntagged} for the 2016, 2017, and 2018 data, respectively.  
The
one- and two-standard deviation ($\sigma$) bands
include only the uncertainty
in the background model normalization associated with the statistical uncertainties of the fits, and are thus shown for illustration purposes only. These bands are obtained using an extended
likelihood fit parametrized in terms of the background yield in bins corresponding to those used in Figs.~\ref{bkgModelFinal2016},~\ref{bkgModelFinal}, and~\ref{fig:statAnalysisBkgValidationPlotsUntagged}.  The corresponding signal model for $\mH=90\GeV$, multiplied by 10, is also shown for illustration purposes.

\section{Systematic uncertainties}

Many of the systematic uncertainties relevant
to the analyses performed in Refs.~\cite{Chatrchyan:2012xdj,HIG-13-001,CMS-PAS-HIG-16-040}
also apply to
this analysis and are described briefly below.
Additional
uncertainties specific to this analysis are described in more detail.

\subsection{Uncertainties evaluated at the per-photon level}

The systematic uncertainties in the
shape of the photon identification BDT distribution and in the
per-photon energy resolution described in Refs.~\cite{HIG-13-001,CMS-PAS-HIG-16-040}
are
applied in this analysis.
These uncertainties
propagate to the multivariate event classifier value, giving rise to the migration of events from one class to another,
and to variations in the per-event efficiency in each class and for each production process. The uncertainties
are evaluated using a signal sample with $\mH=90\GeV$, the midpoint of the search range considered.  For the 2016 data, the largest variation in efficiency due to the photon identification BDT distribution shape is 6.9\%, for the $\ttbar\PH$ process in event class~2; for the 2017 data, 6.4\%, for the $\Pg\Pg\PH$ process in event class~1, and for the 2018 data, 4.1\% 
for the VBF process, also in event class~1. Otherwise, the variations are below 2.2/5.3/3.7\%
for the 2016/2017/2018 data sets.  The largest variation in the efficiency due to
the per-photon energy resolution
applicable to the 2016 data is 18.2\% for the $\Pg\Pg\PH$ process in event class~1; for the 
2017 data, 14.6\%
for the
$\Pg\Pg\PH$ process in class 1, and for the 2018 data, 13.7\%
for the $\ttbar\PH$ process in class 2; otherwise the
variations are below 12.9/13.7/11.9\%
for the 2016/2017/2018 data sets.  

For the 2016/2017/2018 data, uncertainties in the trigger efficiencies give rise to
efficiency variations of less than 0.8/1.8/0.5\%.  Uncertainties in the scale factors of the preselection give rise to efficiency variations of 
less than 3.3/3.2/5.2\% in the case of the 2016/2017/2018 data.  The uncertainties in the scale factors of the electron veto, of the requirement that a photon candidate may not be reconstructed as an electron, and of the minimum value of the photon identification BDT all contribute efficiency variations of less than 1\% for the data of all three years. 

The uncertainties in the measurement and in the correction of the photon energy scale in data, and in the correction of the energy resolution in simulation, arising from the methodology exploiting $\PZ\to\EE$  events as described in Section~\ref{sec:signal} and Refs.~\cite{HIG-13-001,CMS-PAS-HIG-16-040}, are
calculated in the same bins as the corrections themselves.
Uncertainties arising from
modeling of the material budget and of nonuniformity of light collection (the fraction of crystal scintillation light detected as a function of its longitudinal depth when emitted), nonlinearity in the photon energy scale between data and simulation, imperfect electromagnetic shower simulation, and vertex finding~\cite{HIG-13-001,CMS-PAS-HIG-16-040},
are propagated to the parametric signal model, where they result in uncertainties in the diphoton efficiency, mass scale, and resolution.

\subsection{Uncertainties evaluated at the per-event level}

The per-event systematic uncertainty in the total integrated luminosity, estimated from data~\cite{CMS-LUM-17-003,CMSlumi2017,CMSlumi2018}, contributes an uncertainty of 1.2/2.3/2.5\% in the signal yield for the 2016/2017/2018 data.  The overall uncertainty for the 2016--2018 period is 1.6\%.  The uncertainties for each data set are partially correlated to account for common sources in the luminosity measurement schemes. 

For the analysis of the 2016 and 2017 data, an additional uncertainty in the signal efficiency is introduced to account for the gradual shift in the timing of the trigger inputs mentioned previously. It is largest for the VBF class in 2017, and amounts to less than 0.5\,(1.9)\% for 2016\,(2017).

The
systematic uncertainties from the theoretical predictions
considered in this analysis are of three types.  Firstly,
the
uncertainties in the signal acceptance due to changes in particle  \pt and $\eta$ values, arising from variations in the
PDF and
renormalization and factorization scales,
are calculated~\cite{HIG-13-001,CMS-PAS-HIG-16-040}
using a signal sample with $\mH=90\GeV$.  The NNPDF3.0 
(NNPDF3.1) PDF set using the \textsc{MC2hessian} procedure~\cite{Carrazza2015,Gao_2014} is used to estimate the PDF variations for 2016\,(2017 and 2018).  The effects due to variations of the strong coupling strength, \alpS, are also considered, following the
PDF4LHC prescription~\cite{deFlorian:2016spz,Butterworth:2015oua}.  The uncertainties due to
PDF variations are negligible for all three data-taking years.
The largest uncertainty due to
scale variations
in the 2016\,(2017) data is 11.9\,(10.0)\% for the 
$\Pg\Pg\PH$ production process in event class~0.  In the 2018 data, the largest
uncertainties also occur for the
$\Pg\Pg\PH$ process, with the maximum of 13.7\% occurring in the VBF class.
The uncertainties due to variations in
\alpS are largest for $\Pg\Pg\PH$ production, in event class~0 at 1.9\% for the 2016 data, and in the VBF class at 
3.9\,(4.1)\% for the 2017 (2018) data; otherwise, they are below 0.5\%.

Secondly, uncertainties in signal yields due to underlying event and parton shower modeling are obtained using dedicated simulated samples for which the choice and specific tune of the event generator have been modified~\cite{HIG-19-015}. For the underlying event, these uncertainties are largest for the $\ttbar\PH$ production process in event class~2, amounting to 13.7\% for 2016, and for the ZH production process in the VBF event class, amounting to 26.8\,(20.4)\% for 2017 (2018). In the case of the parton shower modeling, they are 
again largest for $\ttbar\PH$ production in event class~2 at 9.6\% for the 2016 data, for $\PZ\PH$  production in the VBF class at 15.9\% for the 2017 data,  and for the VBF production process in the VBF class at 12.0\% for the 2018 data.  Otherwise, they are typically below 7\% for 2016 and  
10\% for 2017 and 2018.

Finally, the uncertainties in the production cross section for
an SM-like Higgs boson,
at a center-of-mass energy of 13\TeV,
are accounted for by 
following
the recommendations of the LHC Higgs cross section working group~\cite{deFlorian:2016spz}.
These uncertainties arise from uncertainties associated with particular choices of PDFs and QCD model parameters.
They are used in the calculation of the expected and observed limits on the product of the
production
cross section and
branching
fraction into two photons relative to the
expected value for an SM-like Higgs boson,
and in the calculations of the expected and observed local $p$-values.
The uncertainty in the branching
fraction into two photons
is neglected.

Uncertainties in the scale factor of the requirement on $\ln(\Sigma\pt^{2}/\GeV^2)$ give rise to efficiency variations of 0.4/0.9/0.5\% in the case of the 2016/2017/2018 data.

For the analysis of the 2017 and 2018 data sets, changes in the presence or properties of jets affect the classification of events, potentially causing them to migrate between the VBF class and the other classes. Jet energy scale corrections account for a maximum uncertainty in signal yields of 14.9\% (2017) and 15.8\% (2018) in the VBF class. For the 2018 data, an additional uncertainty in jet energy scale is applied in order to cover a potential bias in the choice of the leading jets, stemming from a temporary readout interruption of the HCAL region extending between $-3.0<\eta<-1.3$ and $-1.57<\phi<0.87$. This uncertainty is greatest for the VBF event class, ranging between 1 and 3\% with a maximum of 3.4\% for the $\Pg\Pg\PH$ production process. For the other classes, it is in general less than 0.5\%, reaching a maximum of 1.2\% for the VBF production process in event class~0. For the VBF production process, another specific uncertainty in signal event yield arising from the implementation of the matching between matrix element and parton showers in the MC simulation, affecting jet kinematics, is applied; it ranges from 0.1 to 3\% depending on the event class, reaching a maximum of 2.9\,(2.6)\% for the 2017 (2018) data in the VBF event class. The jet energy resolution contributes a signal yield uncertainty of less than 1.1\% in 2017 and 3.0\% in 2018. Finally, an uncertainty in signal yield due to pileup jet identification is estimated by comparing in data and simulation the identification score of jets in events with a \PZ boson and one balanced jet, and amounts to less than 4.1\% in 2017 and 3.2\% in 2018.

An additional source of per-event systematic uncertainty specific to this analysis is the modeling of the \PZ boson resonance component
of the background.
As explained previously, the parameters of the DCB function used to model the \PZ boson resonance are obtained from
doubly-misidentified events, which are simulated DY events with all selection requirements applied including the electron veto requirements.
These parameters could be different
for data and simulation.  To estimate these differences, we study simulated singly-misidentified events from the DY, diphoton,
\GAMJET, and QCD multijet processes.  We fit the invariant diphoton mass of these events in data, in
simulation including the sum of all background processes,
and in simulated DY events alone,
with a DCB + 
exponential component
that describes the additional continuum background inherent in singly-misidentified events. We consider the pairwise differences
among the DCB
mean and standard deviation parameters extracted from these three types of fits for each event class.
The differences are considered
statistically significant if they are greater than the
quadratic sum of the statistical
uncertainties from the fit.
These differences contribute to the total systematic
uncertainty in the
DCB parameter values.
The nominal parameter values are obtained from
doubly-misidentified events so
the
differences contributing to the parameter uncertainties that are estimated from singly-misidentified events are doubled, to reflect the
more conservative case where the parameters of the two
photon candidates in a
doubly-misidentified event
are completely correlated.

{\tolerance=800 The total systematic
uncertainty in each event class
for
the
mean and standard deviation parameters
is the quadratic sum of:  the statistical
uncertainty from the fit
to the
doubly-misidentified simulated DY events;
the
doubled difference
between the  parameter values
from data and from the sum of all simulated background processes, and the doubled difference between the parameter values from
the sum of all simulated background processes
and from simulated DY events alone,
determined from the singly-misidentified events.  
In the case of the VBF class, an additional term, the doubled difference between the parameter values from simulated singly-misidentified DY events with and without the requirement on the minimum value of the combined BDT classifier, is added in the quadratic sum.
\par}

An uncertainty is assigned to the fixed values of the DCB parameters $\alpha_{\text{L}}$ and $\alpha_{\text{R}}$ for each event class by performing the fits of the doubly-misidentified simulated $\PZ\to\EE$ events with these two parameters allowed to float. The uncertainty is the quadratic sum of the statistical uncertainty in the fitted values of $\alpha_{\text{L}}$ and $\alpha_{\text{R}}$, and the systematic uncertainty, which is taken as the difference between the fixed and fitted values for each of the two parameters.

Finally, the analysis takes into account the statistical uncertainties in the values of the DCB $n_{\text{L}}$ and $n_{\text{R}}$ parameters, and in the values of the additional exponential function $p$ and $g$ parameters, obtained from the
fits to the doubly-misidentified simulated $\PZ\to\EE$ events.

The invariant mass distribution of electron pairs from surviving DY events has a shape that is similar to that of photon pairs in signal, when \mH is close to the fitted peak of the doubly-misidentified events.  The systematic uncertainty in the fraction, $f$, of background events attributed to the surviving DY component, has been adjusted to minimize the potential bias induced by that shape similarity, while at the same time yielding a negligible effect on the expected limits and expected significances.  

\section{Results}

Table~\ref{tab:statAnalysisSigBkgYields} shows the expected numbers of signal events corresponding to the
production of a hypothetical
additional SM-like
Higgs boson with $\mH=90\GeV$, from the
analyses of the 2016, 2017, and 2018 data.  The total number is broken down
into the
contributions
from all the production processes
in each of the event classes, where the $\mathrm{V}\PH$ processes ($\PW\PH$
and $\PZ\PH$) are listed separately.  Also shown are the $\sigma_{\text{eff}}$ and $\sigma_{\text{HM}}$ (defined as the FWHM divided by 2.35) values, as well as the
number of background events per \GeV estimated from the background-only fit to the data, that includes the number, shown separately, 
from the DY process,
in the corresponding $\sigma_{\text{eff}}$ window centered on $\mH=90\GeV$, using the chosen background function.
\begin{table*}[!tb]
\centering
\topcaption{The expected number of SM-like Higgs boson signal events ($\mH=90\GeV$) per event class and the corresponding percentage breakdown per production process, for the 2016, 2017, and 2018 data. The values of $\sigma_{\text{eff}}$ and $\sigma_{\text{HM}}$
are also shown,
along with the
number of background events (``Bkg.'') per \GeV estimated from the background-only fit to the data, that includes the number, shown separately, 
from the DY process (``DY Bkg.''), in a $\sigma_{\text{eff}}$ window centered on $\mH=90\GeV$.
}
\cmsTable{
\begin{tabular}{lrrrrrrrccrr}
\hline
\multicolumn{2}{l}{\multirow{3}{*}{Event classes}} & \multicolumn{8}{c}{Expected SM-like Higgs boson signal yield ($\mH=90\GeV$)} & \multicolumn{1}{c}{Bkg.} & \multicolumn{1}{c}{DY Bkg.}  \\
\cline{3-10}
\multicolumn{2}{c}{} & Total & $\Pg\Pg\PH$ & VBF & $\PW\PH$ & $\PZ\PH$
& $\ttbar\PH$ & $\sigma_\text{eff}$ & $\sigma_\mathrm{HM}$ & \multicolumn{1}{c}{(${\GeVns}^{-1}$)} & \multicolumn{1}{c}{(${\GeVns}^{-1}$)}  \\
\multicolumn{2}{c}{} & & (\%) & (\%) & (\%) & (\%) & (\%) & (\GeVns) & (\GeVns) &  & \\
\hline
2016 & 0 &    130  &  71.9  &  15.6  &  6.2  &  3.6  &  2.6 & 1.12 & 1.00 & 271 & 12 \\
36.3\fbinv & 1 &    304  &  87.4  &  6.6  &  3.6  &  2.1  &  0.3 & 1.25 & 1.07 & 3093 & 33 \\
& 2 &    407  &  94.7  &  2.5  &  1.7  &  1.0  &  0.1 & 1.87 & 1.51 & 9190 & 193 \\
&Total &    842  &  88.5  &  6.0  &  3.1  &  1.8  &  0.6 & 1.50 & 1.20 & 12\,554  & 239 \\
[\cmsTabSkip]
2017 & 0 &    104  &  73.4  &  11.6  &  7.5  &  4.3  &  3.2 & 1.27 & 1.13 & 248 & 7 \\
41.5\fbinv & 1 &    347  &  88.5  &  5.6  &  3.5  &  2.1  &  0.3 & 1.40 & 1.24 & 3625 & 83 \\
& 2 &    413  &  94.4  &  2.6  &  1.9  &  1.1  &  0.1 & 1.91 & 1.64 & 8169 & 244 \\
& VBF &    26  &  45.6  &  51.8  &  1.0  &  0.5  &  1.0 & 1.33 & 1.15 & 29 & 1 \\
&Total &    890  &  88.2  &  6.2  &  3.1  &  1.8  &  0.6 & 1.60 & 1.35 & 12\,071  & 338 \\
[\cmsTabSkip]
2018 &0 &    162  &  75.1  &  10.2  &  7.3  &  4.3  &  3.0 & 1.21 & 1.05 & 430 & 3 \\
54.4\fbinv & 1 &    585  &  90.1  &  4.8  &  3.1  &  1.8  &  0.2 & 1.34 & 1.17 & 6445 & 378 \\
& 2 &    473  &  94.4  &  2.5  &  1.9  &  1.2  &  0.1 & 2.01 & 1.73 & 10\,982 & 720 \\
& VBF &    38  &  45.4  &  51.9  &  1.1  &  0.6  &  1.0 & 1.21 & 1.03 & 46 & 1 \\
&Total &    1258  &  88.4  &  6.1  &  3.1  &  1.8 &  0.6 & 1.54 & 1.27 & 17\,902 & 1104 \\
\hline
\end{tabular}
}

\label{tab:statAnalysisSigBkgYields}
\end{table*}

\begin{figure}[!htbp]\centering
   \includegraphics[width=0.49\textwidth]{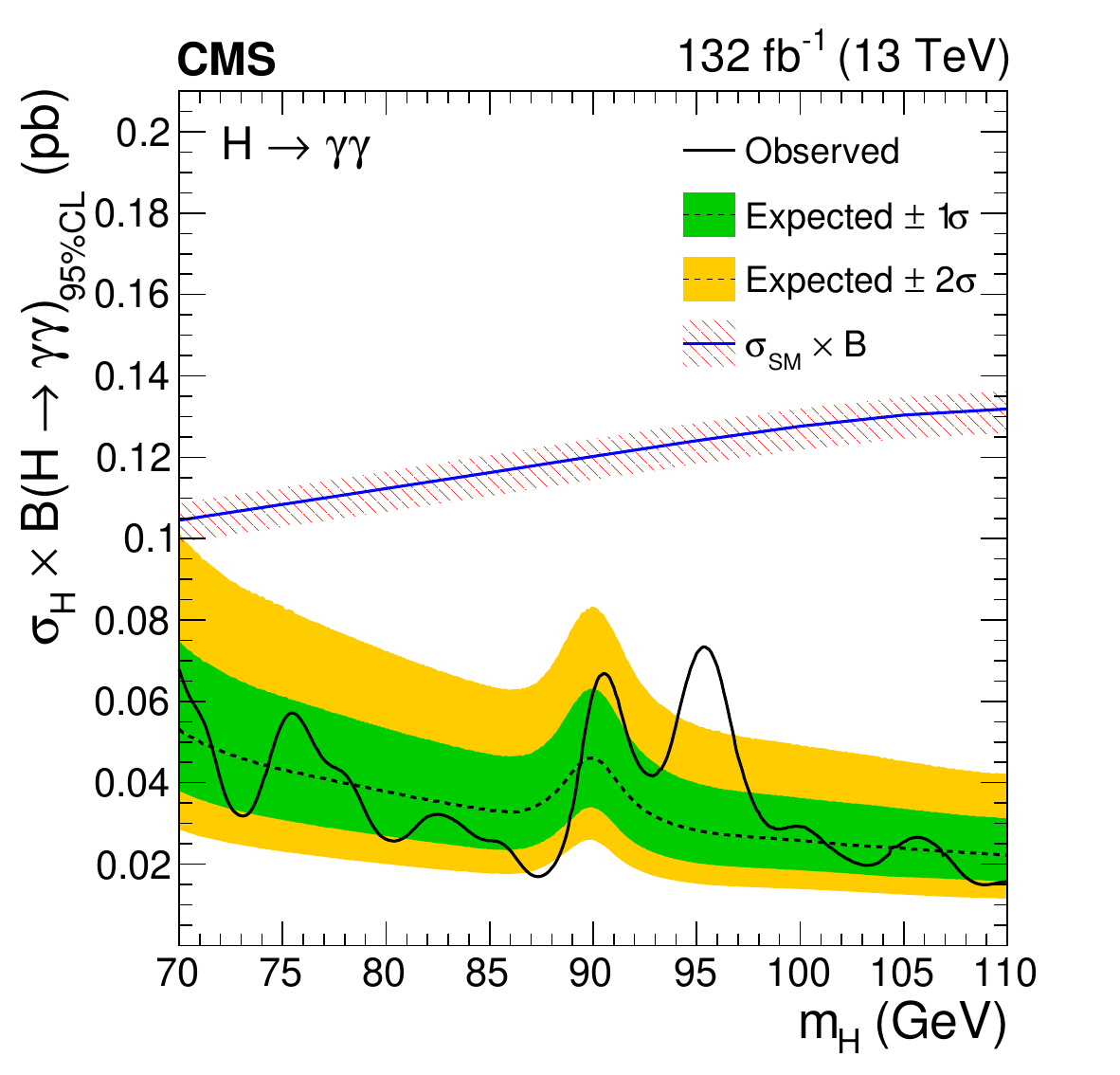}
    \includegraphics[width=0.49\textwidth]{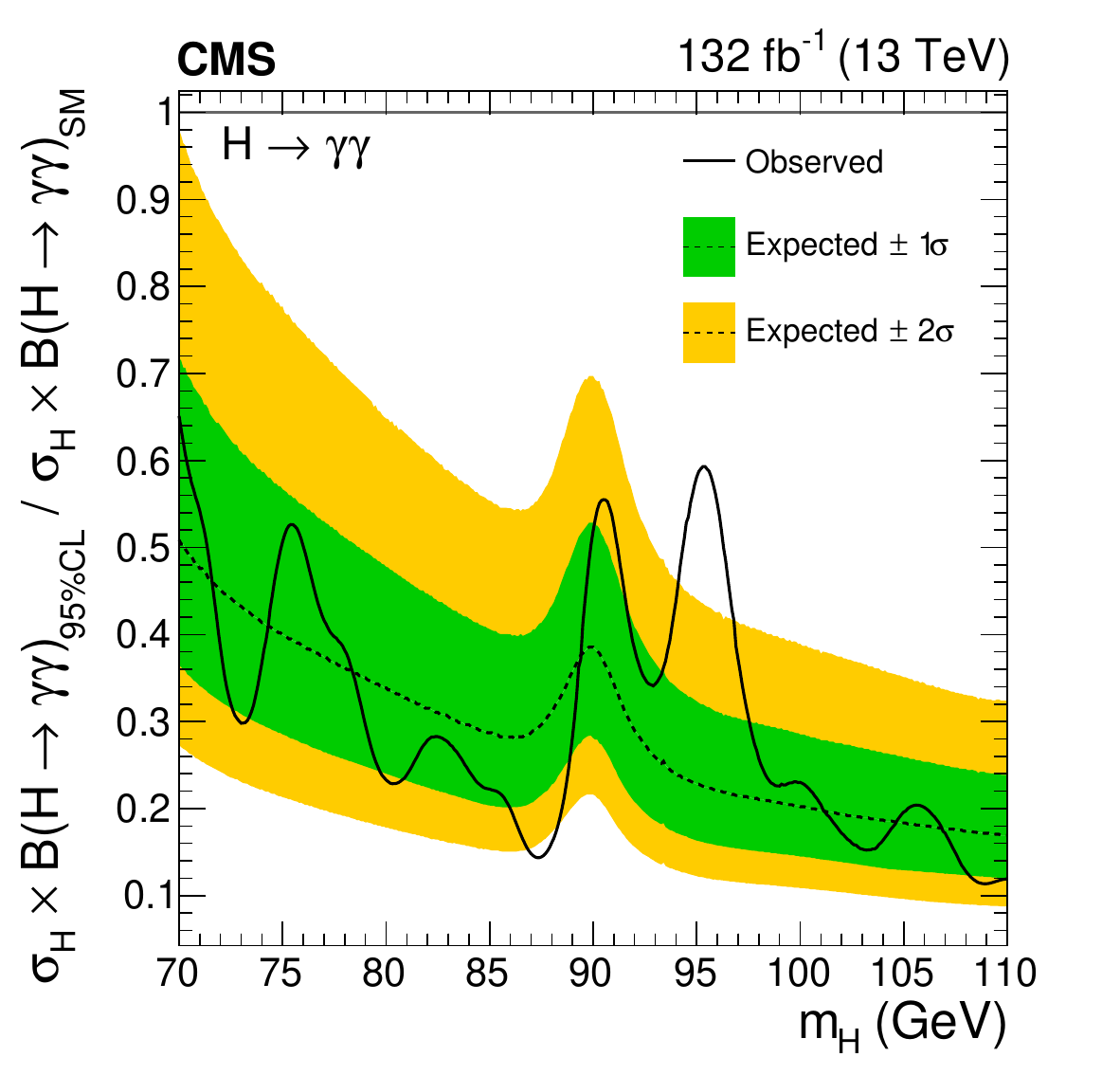}
   \caption{Expected and observed exclusion limits (95\% \CL, in the asymptotic approximation) on the product of the production cross section and branching fraction into two photons for an additional SM-like Higgs boson (\cmsLeft); the theoretical prediction for the product of the cross section and branching fraction into two photons for an additional SM-like Higgs boson is shown as a solid line with
a hatched band, indicating its uncertainty~\cite{deFlorian:2016spz}.  The same limits shown relative to the expected SM-like value (\cmsRight).  The results correspond to the combined 2016, 2017, and 2018 data sets.  The inner and outer bands indicate the regions containing the distribution of limits located within ${\pm}1\sigma$ and ${\pm}2\sigma$,
respectively, of the expectation under the background-only hypothesis.}\label{FullRun2LimitPvalue}
\end{figure}

A simultaneous binned maximum likelihood fit to the diphoton invariant mass distributions in all event classes and in all three data-taking years,
with a step size of 0.1\GeV,
 is performed over the range $70<\mgg<110\GeV$.  The upper limits on the product of the cross section $\sigma_{\PH}$  and branching fraction $\cal{B}$ into two photons are set using
the \CLs modified frequentist criterion~\cite{LHC-HCG,Junk:1999kv,Read:2000ru}.  This construction uses the profile-likelihood ratio as the test statistic~\cite{Cowan:2010js} under the asymptotic approximation.  All of the experimental systematic uncertainties, except those pertaining to the integrated luminosity as previously noted, are assumed to be uncorrelated among the different data sets,
as are the theoretical uncertainties
in the signal acceptance due to PDF variations.  The theoretical uncertainties
in the signal acceptance due to 
scale variations as well as 
in the production cross sections at the center-of-mass energy of 13\TeV for an
additional SM-like
Higgs boson are assumed to be
fully correlated.
Figure~\ref{FullRun2LimitPvalue} shows the expected 
and observed 
95\% confidence level (\CL) upper limits on the product of  $\sigma_{\PH}$ and $\cal{B}$ into two photons, as a function of the mass of an additional SM-like Higgs boson in the range $70<\mgg<110\GeV$, calculated with respect to the background-only hypothesis, from the statistical combination of the 2016, 2017, and 2018 data sets.

The minimum (maximum) observed upper limit on the product of the production cross section
and branching fraction is approximately 15\,(73)\unit{fb}, corresponding to a mass hypothesis of 108.9\,(95.4)\GeV.

\begin{figure*}[!htb]\centering
\includegraphics[width=0.49\textwidth]{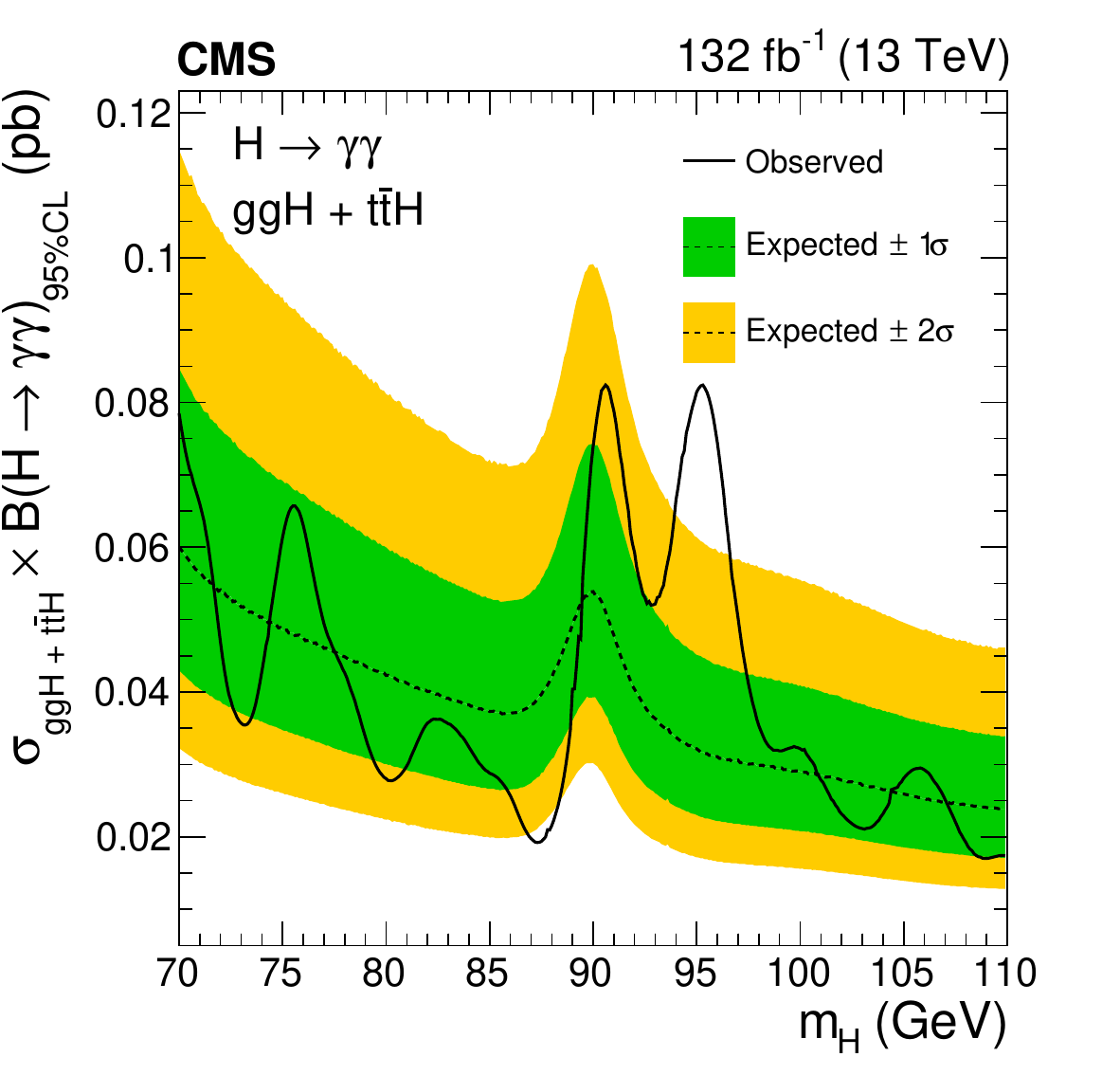}
\includegraphics[width=0.49\textwidth]{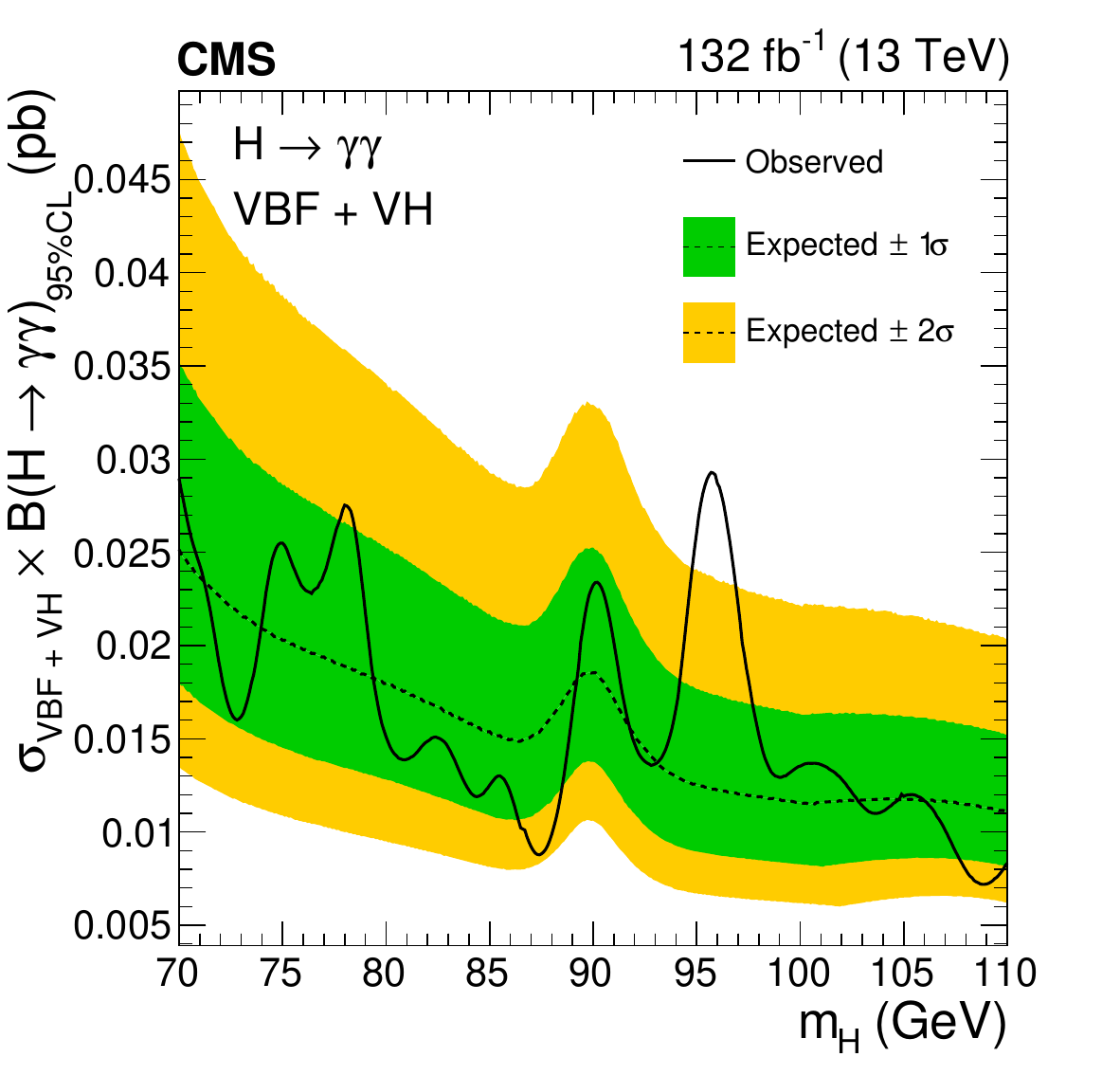} \\
\includegraphics[width=0.49\textwidth]{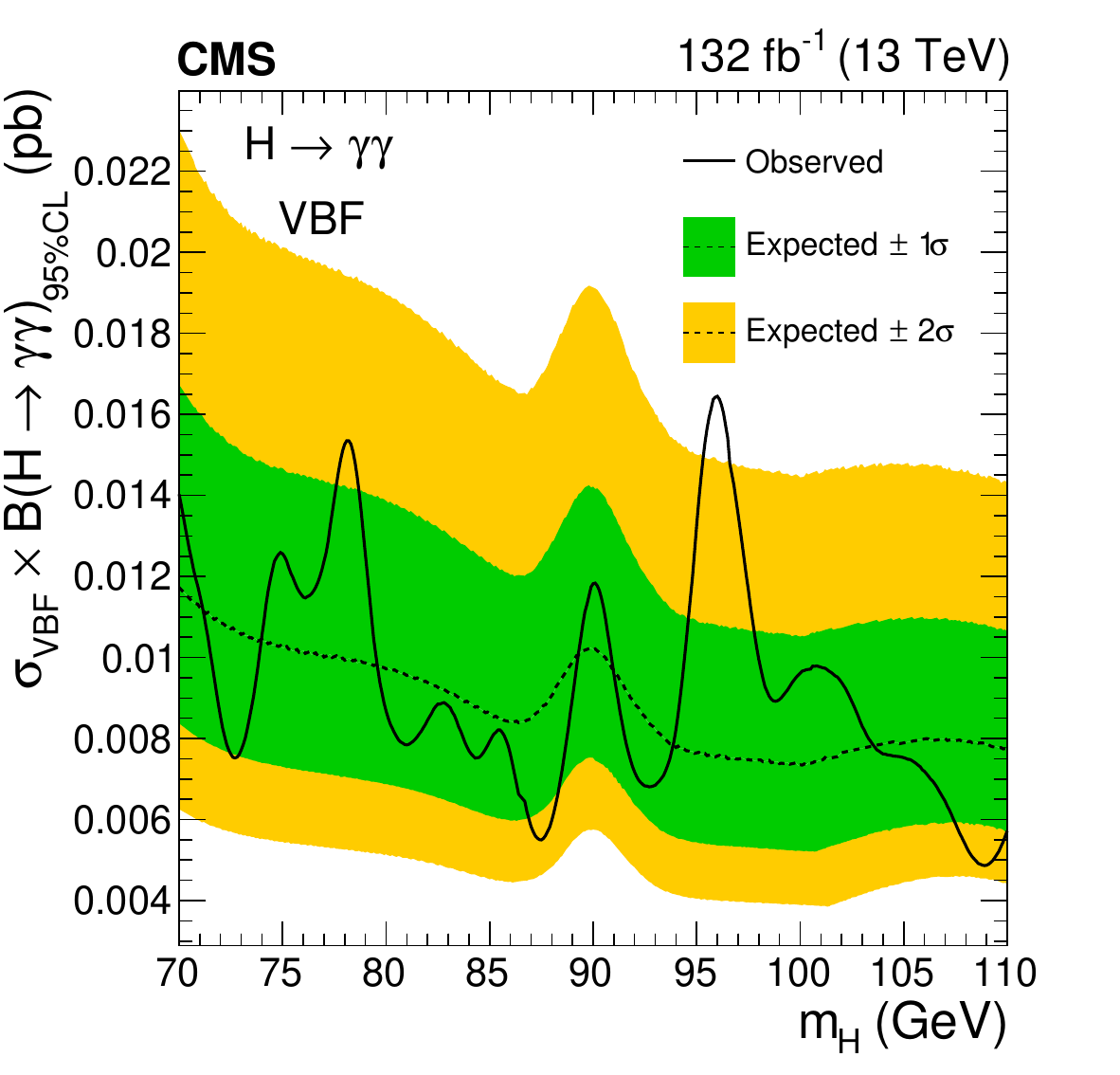}
\includegraphics[width=0.49\textwidth]{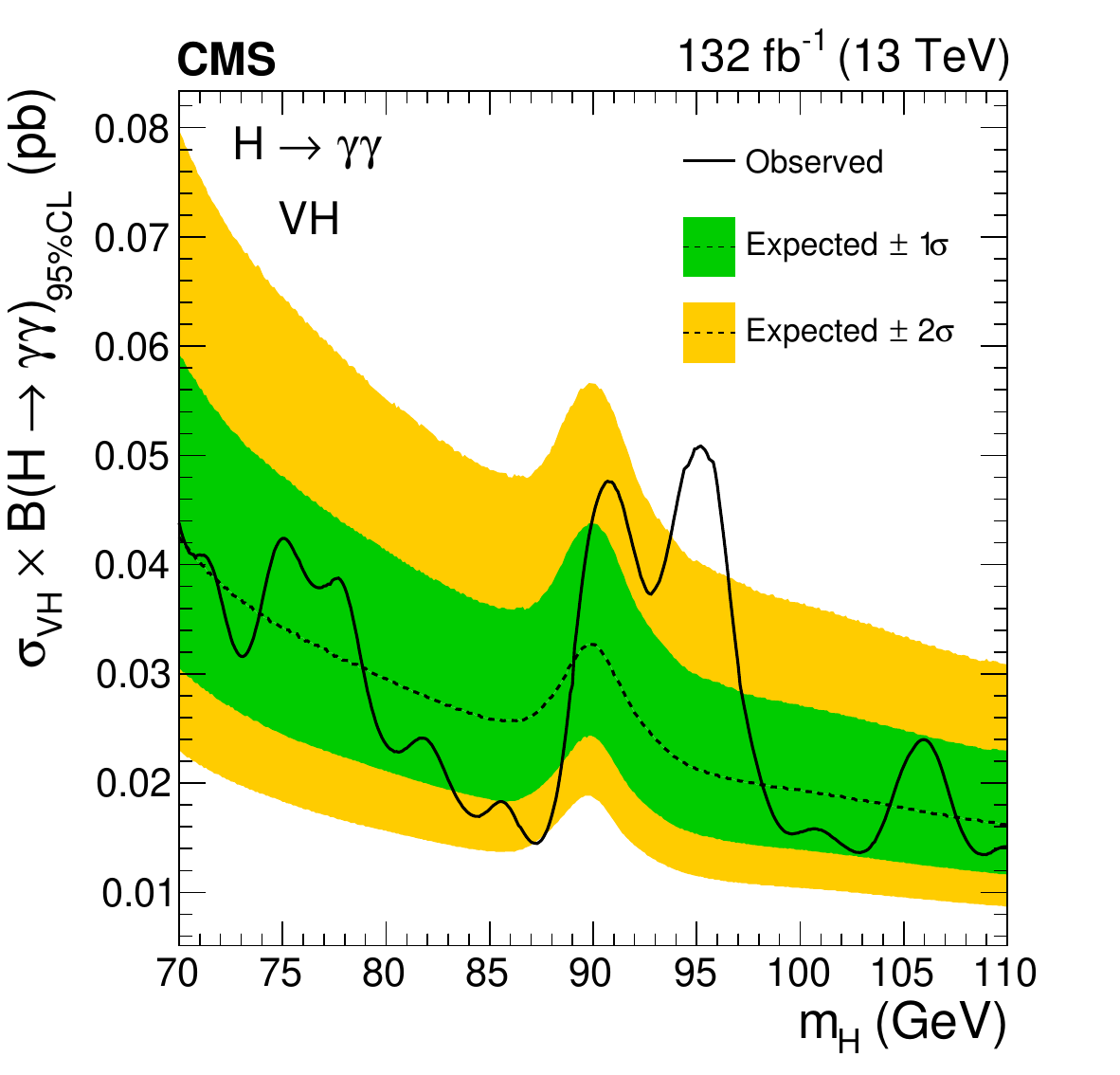}
\caption{Expected and observed exclusion limits (95\% \CL, in the asymptotic approximation) on the product of the production cross section and branching
fraction into two photons for an additional SM-like Higgs boson, for the $\Pg\Pg\PH$ plus $\ttbar\PH$ (upper left) and VBF plus $\mathrm{V}\PH$ (upper right) processes, and assuming 100\% production via the VBF (lower left) or $\mathrm{V}\PH$ (lower right) processes, from the statistical combination of the 2016, 2017, and 2018 data sets.  The inner and outer bands indicate the regions containing the distribution of limits located within $\pm 1\sigma$ and $\pm 2\sigma$, respectively, of the expectation under the background-only hypothesis.}
\label{fig:8TeVLimitSplitAbs}
\end{figure*}

In addition, the expected and observed 95\% \CL upper limits for the production of an additional SM-like Higgs boson via the sum of the fermion-coupled $\Pg\Pg\PH$ and $\ttbar\PH$ processes are shown in Fig.~\ref{fig:8TeVLimitSplitAbs}, for the combined data set. The limits for production via the sum of the vector boson-coupled VBF and $\mathrm{V}\PH$ processes are also shown.  The production processes, in each case, are combined assuming relative proportions as predicted by the SM. The same limits are shown under the assumption of 100\% production via each of the VBF and $\mathrm{V}\PH$ processes.

\begin{figure}[htb]\centering
\includegraphics[width=\cmsFigWidth]{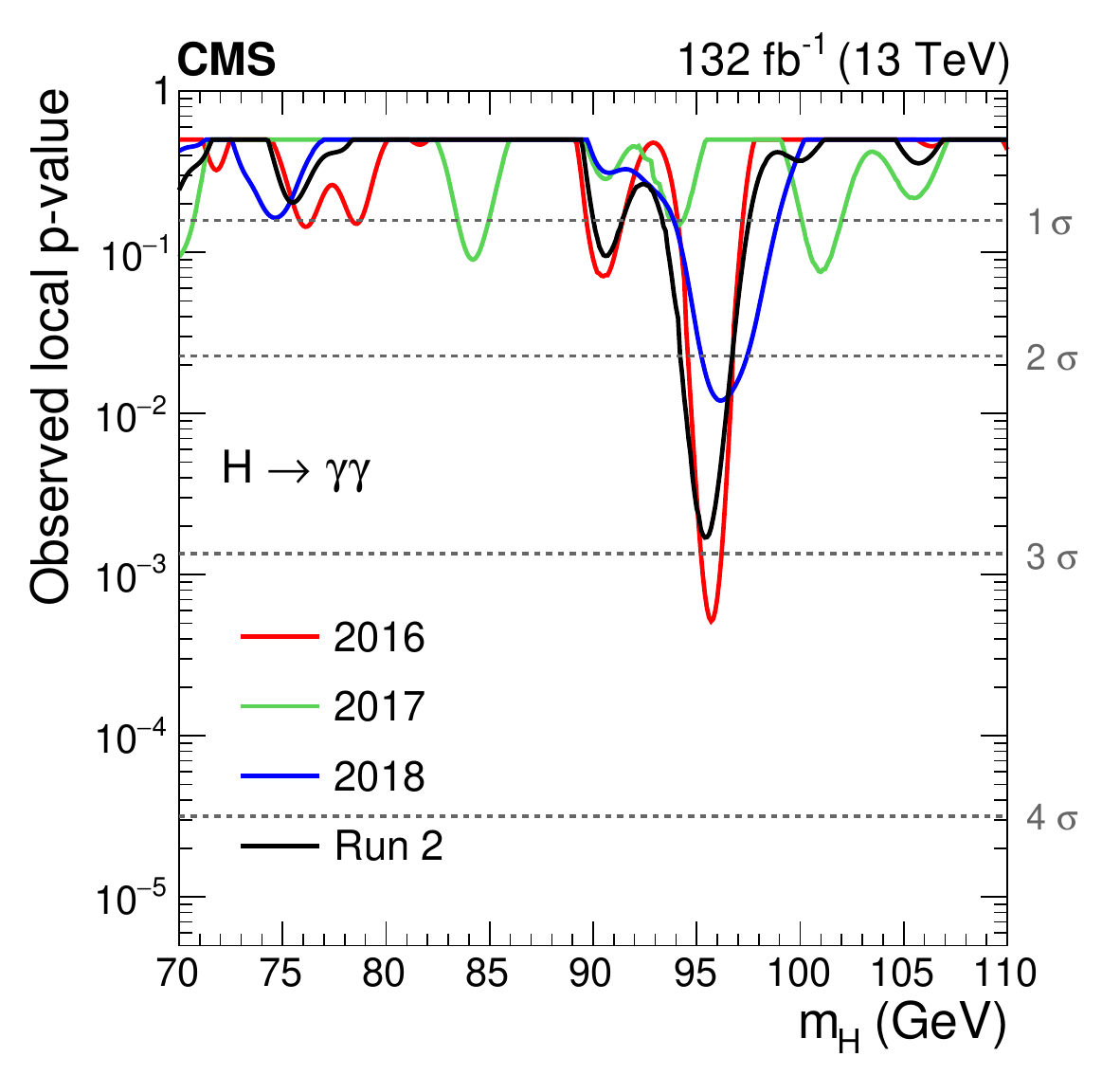}
\caption{The observed local $p$-values for an additional SM-like Higgs boson as a function of \mH, from the analysis of the data from 2016, 2017, 2018, and 
their combination.}
\label{fig:decomposedPvalue}
\end{figure}

Figure~\ref{fig:decomposedPvalue} shows the observed 
local $p$-values as a function of the mass of an
additional SM-like
Higgs boson, calculated with respect to the background-only hypothesis,
from the analyses of the data from 2016, 2017, 2018, and their statistical combination.  The 2016 data reanalyzed with the new detector calibration show an excess that has increased in local significance with respect to the previous analysis~\cite{HIG-17-013} from 2.8$\sigma$ to 3.3$\sigma$ at a mass of approximately 95.4\GeV.  The local (global) significance decreases to  2.9$\sigma$\, (1.3$\sigma$) when the data from 2017 and 2018 are included.  The global significance has been calculated
using the method of Ref.~\cite{pvalue}.

\section{Summary}

A search for an additional, SM-like, low-mass Higgs boson decaying into two photons has been presented. It is based upon data samples corresponding to an integrated luminosity of 132\fbinv collected in $\Pp\Pp$ collisions at a center-of-mass energy of 13\TeV in 2016--2018.  The search is performed in a mass range between 70 and 110\GeV. The expected and observed 95\% \CL upper limits on the product of the production cross section and branching fraction into two photons for an additional SM-like Higgs boson as well as the expected and observed local $p$-values are presented.  The observed upper limit on the product of the production cross section and branching fraction for the full data set ranges from 15 to 73\unit{fb}. The results of the statistical combination of the analyses of the three data sets show no significant excess over the background expectation. The maximum deviation with respect to the background is seen for a mass hypothesis of 95.4\GeV with a local (global) significance of 2.9\,(1.3) standard deviations.

\begin{acknowledgments}
We congratulate our colleagues in the CERN accelerator departments for the excellent performance of the LHC and thank the technical and administrative staffs at CERN and at other CMS institutes for their contributions to the success of the CMS effort. In addition, we gratefully acknowledge the computing centers and personnel of the Worldwide LHC Computing Grid and other centers for delivering so effectively the computing infrastructure essential to our analyses. Finally, we acknowledge the enduring support for the construction and operation of the LHC, the CMS detector, and the supporting computing infrastructure provided by the following funding agencies: SC (Armenia), BMBWF and FWF (Austria); FNRS and FWO (Belgium); CNPq, CAPES, FAPERJ, FAPERGS, and FAPESP (Brazil); MES and BNSF (Bulgaria); CERN; CAS, MoST, and NSFC (China); MINCIENCIAS (Colombia); MSES and CSF (Croatia); RIF (Cyprus); SENESCYT (Ecuador); ERC PRG, RVTT3 and MoER TK202 (Estonia); Academy of Finland, MEC, and HIP (Finland); CEA and CNRS/IN2P3 (France); SRNSF (Georgia); BMBF, DFG, and HGF (Germany); GSRI (Greece); NKFIH (Hungary); DAE and DST (India); IPM (Iran); SFI (Ireland); INFN (Italy); MSIP and NRF (Republic of Korea); MES (Latvia); LMTLT (Lithuania); MOE and UM (Malaysia); BUAP, CINVESTAV, CONACYT, LNS, SEP, and UASLP-FAI (Mexico); MOS (Montenegro); MBIE (New Zealand); PAEC (Pakistan); MES and NSC (Poland); FCT (Portugal); MESTD (Serbia); MCIN/AEI and PCTI (Spain); MOSTR (Sri Lanka); Swiss Funding Agencies (Switzerland); MST (Taipei); MHESI and NSTDA (Thailand); TUBITAK and TENMAK (Turkey); NASU (Ukraine); STFC (United Kingdom); DOE and NSF (USA).

\hyphenation{Rachada-pisek} Individuals have received support from the Indo-French Network in High Energy Physics financed by the Indo-French Center for the Promotion of Advanced Research (CEFIPRA/IFCPAR); the Marie-Curie program and the European Research Council and Horizon 2020 Grant, contract Nos.\ 675440, 724704, 752730, 758316, 765710, 824093, 101115353, 101002207, and COST Action CA16108 (European Union); the Leventis Foundation; the Alfred P.\ Sloan Foundation; the Alexander von Humboldt Foundation; the Science Committee, project no. 22rl-037 (Armenia); the Belgian Federal Science Policy Office; the Fonds pour la Formation \`a la Recherche dans l'Industrie et dans l'Agriculture (FRIA-Belgium); the Agentschap voor Innovatie door Wetenschap en Technologie (IWT-Belgium); the F.R.S.-FNRS and FWO (Belgium) under the ``Excellence of Science -- EOS" -- be.h project n.\ 30820817; the Beijing Municipal Science \& Technology Commission, No. Z191100007219010 and Fundamental Research Funds for the Central Universities (China); the Ministry of Education, Youth and Sports (MEYS) of the Czech Republic; the Shota Rustaveli National Science Foundation, grant FR-22-985 (Georgia); the Deutsche Forschungsgemeinschaft (DFG), under Germany's Excellence Strategy -- EXC 2121 ``Quantum Universe" -- 390833306, and under project number 400140256 - GRK2497; the Hellenic Foundation for Research and Innovation (HFRI), Project Number 2288 (Greece); the Hungarian Academy of Sciences, the New National Excellence Program - \'UNKP, the NKFIH research grants K 131991, K 133046, K 138136, K 143460, K 143477, K 146913, K 146914, K 147048, 2020-2.2.1-ED-2021-00181, and TKP2021-NKTA-64 (Hungary); the Council of Science and Industrial Research, India; ICSC -- National Research Center for High Performance Computing, Big Data and Quantum Computing and FAIR -- Future Artificial Intelligence Research, funded by the NextGenerationEU program (Italy); the Latvian Council of Science; the Ministry of Education and Science, project no. 2022/WK/14, and the National Science Center, contracts Opus 2021/41/B/ST2/01369 and 2021/43/B/ST2/01552 (Poland); the Funda\c{c}\~ao para a Ci\^encia e a Tecnologia, grant CEECIND/01334/2018 (Portugal); the National Priorities Research Program by Qatar National Research Fund; MCIN/AEI/10.13039/501100011033, ERDF ``a way of making Europe", and the Programa Estatal de Fomento de la Investigaci{\'o}n Cient{\'i}fica y T{\'e}cnica de Excelencia Mar\'{\i}a de Maeztu, grant MDM-2017-0765 and Programa Severo Ochoa del Principado de Asturias (Spain); the Chulalongkorn Academic into Its 2nd Century Project Advancement Project, and the National Science, Research and Innovation Fund via the Program Management Unit for Human Resources \& Institutional Development, Research and Innovation, grant B37G660013 (Thailand); the Kavli Foundation; the Nvidia Corporation; the SuperMicro Corporation; the Welch Foundation, contract C-1845; and the Weston Havens Foundation (USA).
\end{acknowledgments}

\bibliography{auto_generated}

\providecommand{\href}[2]{#2}\begingroup\raggedright\begin{thebibliography}{10}%
\makeatletter
\providecommand{\hrefCMSnoop }[0]{\@secondoftwo}%
\makeatother
\providecommand{\doi}{\texttt{doi:}\begingroup \urlstyle{tt}\Url}

\bibitem{SMHiggsO}
\hrefCMSnoop {}{S.~L. Glashow, ``Partial-symmetries of weak interactions'',}
  \textit{ Nucl. Phys.} \textbf{ 22} (1961) 579,
\href{http://dx.doi.org/10.1016/0029-5582(61)90469-2}{\doi{10.1016/0029-5582(61)90469-2}}.

\bibitem{SMHiggsT}
\hrefCMSnoop {}{S.~Weinberg, ``A model of leptons'',} \textit{ Phys. Rev.
  Lett.} \textbf{ 19} (1967) 1264,
\href{http://dx.doi.org/10.1103/PhysRevLett.19.1264}{\doi{10.1103/PhysRevLett.19.1264}}.

\bibitem{SMHiggsTh}
\hrefCMSnoop {}{A.~Salam, ``Weak and electromagnetic interactions'',} in
  \textit{ Elementary particle physics: relativistic groups and analyticity},
  N.~Svartholm, ed., p.~367.
\newblock Almqvist \& Wiksell, Stockholm, 1968.
\newblock Proceedings of the eighth Nobel symposium.

\bibitem{HiggsExpO}
\hrefCMSnoop {}{F.~Englert and R.~Brout, ``Broken symmetry and the mass of
  gauge vector mesons'',} \textit{ Phys. Rev. Lett.} \textbf{ 13} (1964) 321,
  \href{http://dx.doi.org/10.1103/PhysRevLett.13.321}{\doi{10.1103/PhysRevLett.13.321}}.

\bibitem{HiggsExpT}
\hrefCMSnoop {}{P.~W. Higgs, ``Broken symmetries, massless particles and gauge
  fields'',} \textit{ Phys. Lett.} \textbf{ 12} (1964) 132,
\href{http://dx.doi.org/10.1016/0031-9163(64)91136-9}{\doi{10.1016/0031-9163(64)91136-9}}.

\bibitem{HiggsExpTh}
\hrefCMSnoop {}{P.~W. Higgs, ``Broken symmetries and the masses of gauge
  bosons'',} \textit{ Phys. Rev. Lett.} \textbf{ 13} (1964) 508,
\href{http://dx.doi.org/10.1103/PhysRevLett.13.508}{\doi{10.1103/PhysRevLett.13.508}}.

\bibitem{PhysRevLett.13.585}
\hrefCMSnoop {}{G.~S. Guralnik, C.~R. Hagen, and T.~W.~B. Kibble, ``Global
  conservation laws and massless particles'',} \textit{ Phys. Rev. Lett.}
  \textbf{ 13} (1964) 585,
  \href{http://dx.doi.org/10.1103/PhysRevLett.13.585}{\doi{10.1103/PhysRevLett.13.585}}.

\bibitem{PhysRev.145.1156}
\hrefCMSnoop {}{P.~W. Higgs, ``Spontaneous symmetry breakdown without massless
  bosons'',} \textit{ Phys. Rev.} \textbf{ 145} (1966) 1156,
  \href{http://dx.doi.org/10.1103/PhysRev.145.1156}{\doi{10.1103/PhysRev.145.1156}}.

\bibitem{PhysRev.155.1554}
\hrefCMSnoop {}{T.~W.~B. Kibble, ``Symmetry breaking in non-{Abelian} gauge
  theories'',} \textit{ Phys. Rev.} \textbf{ 155} (1967) 1554,
  \href{http://dx.doi.org/10.1103/PhysRev.155.1554}{\doi{10.1103/PhysRev.155.1554}}.

\bibitem{Aad:2012tfa}
\hrefCMSnoop {}{{ATLAS Collaboration}, ``Observation of a new particle in the
  search for the standard model {Higgs} boson with the {ATLAS} detector at the
  {LHC}'',} \textit{ Phys. Lett. B} \textbf{ 716} (2012) 1,
  \href{http://dx.doi.org/10.1016/j.physletb.2012.08.020}{\doi{10.1016/j.physletb.2012.08.020}},
\href{http://www.arXiv.org/abs/1207.7214}{\texttt{arXiv:1207.7214}}.

\bibitem{Chatrchyan:2012xdj}
\hrefCMSnoop {}{{CMS Collaboration}, ``Observation of a new boson at a mass of
  125 {GeV} with the {CMS} experiment at the {LHC}'',} \textit{ Phys. Lett. B}
  \textbf{ 716} (2012) 30,
  \href{http://dx.doi.org/10.1016/j.physletb.2012.08.021}{\doi{10.1016/j.physletb.2012.08.021}},
\href{http://www.arXiv.org/abs/1207.7235}{\texttt{arXiv:1207.7235}}.

\bibitem{Chatrchyan:2013lba}
\hrefCMSnoop {}{{CMS Collaboration}, ``Observation of a new boson with mass
  near 125 {GeV} in {$\Pp\Pp$} collisions at $\sqrt{s}$ = 7 and 8 {TeV}'',}
  \textit{ JHEP} \textbf{ 06} (2013) 081,
  \href{http://dx.doi.org/10.1007/JHEP06(2013)081}{\doi{10.1007/JHEP06(2013)081}},
\href{http://www.arXiv.org/abs/1303.4571}{\texttt{arXiv:1303.4571}}.

\bibitem{ATLASNature}
\hrefCMSnoop {}{{ATLAS Collaboration}, ``A detailed map of {Higgs} boson
  interactions by the {ATLAS} experiment ten years after the discovery'',}
  \textit{ Nature} \textbf{ 607} (2022) 52,
  \href{http://dx.doi.org/10.1038/s41586-022-04893-w}{\doi{10.1038/s41586-022-04893-w}},
\href{http://www.arXiv.org/abs/2207.00092}{\texttt{arXiv:2207.00092}}.

\bibitem{CMSNature}
\hrefCMSnoop {}{{CMS Collaboration}, ``A portrait of the {Higgs} boson by the
  {CMS} experiment ten years after the discovery'',} \textit{ Nature} \textbf{
  607} (2022) 60,
  \href{http://dx.doi.org/10.1038/s41586-022-04892-x}{\doi{10.1038/s41586-022-04892-x}},
\href{http://www.arXiv.org/abs/2207.00043}{\texttt{arXiv:2207.00043}}.

\bibitem{Celis_2013}
\hrefCMSnoop {}{A.~Celis, V.~Ilisie, and A.~Pich, ``{LHC constraints on
  two-Higgs-doublet models}'',} \textit{ JHEP} \textbf{ 07} (2013) 053,
  \href{http://dx.doi.org/10.1007/JHEP07(2013)053}{\doi{10.1007/JHEP07(2013)053}},
  \href{http://www.arXiv.org/abs/1302.4022}{\texttt{arXiv:1302.4022}}.

\bibitem{Coleppa:2013dya}
\hrefCMSnoop {}{B.~Coleppa, F.~Kling, and S.~Su, ``{Constraining type II 2HDM
  in light of LHC Higgs searches}'',} \textit{ JHEP} \textbf{ 01} (2014) 161,
  \href{http://dx.doi.org/10.1007/JHEP01(2014)161}{\doi{10.1007/JHEP01(2014)161}},
  \href{http://www.arXiv.org/abs/1305.0002}{\texttt{arXiv:1305.0002}}.

\bibitem{Chang:2013ona}
S.~Chang\hrefCMSnoop {}{ { et~al.}, ``Two-{Higgs}-doublet models for the {LHC
  Higgs} boson data at $\sqrt{s}=$ 7 and 8 {TeV}'',} \textit{ JHEP} \textbf{
  09} (2014) 101,
  \href{http://dx.doi.org/10.1007/JHEP09(2014)101}{\doi{10.1007/JHEP09(2014)101}},
\href{http://www.arXiv.org/abs/1310.3374}{\texttt{arXiv:1310.3374}}.

\bibitem{Bernon:2015wef}
J.~Bernon\hrefCMSnoop {}{ { et~al.}, ``Scrutinizing the alignment limit in
  two-{Higgs}-doublet models. {II}. {$\mH=125$}{\GeV}'',} \textit{ Phys. Rev.
  D} \textbf{ 93} (2016) 035027,
  \href{http://dx.doi.org/10.1103/PhysRevD.93.035027}{\doi{10.1103/PhysRevD.93.035027}},
\href{http://www.arXiv.org/abs/1511.03682}{\texttt{arXiv:1511.03682}}.

\bibitem{Cacciapaglia2016}
G.~Cacciapaglia\hrefCMSnoop {}{ { et~al.}, ``Search for a lighter {Higgs} boson
  in two-{Higgs}-doublet models'',} \textit{ JHEP} \textbf{ 12} (2016) 68,
  \href{http://dx.doi.org/10.1007/JHEP12(2016)068}{\doi{10.1007/JHEP12(2016)068}},
\href{http://www.arXiv.org/abs/1607.08653}{\texttt{arXiv:1607.08653}}.

\bibitem{Fayet:1974pd}
\hrefCMSnoop {}{P.~Fayet, ``Supergauge invariant extension of the {Higgs}
  mechanism and a model for the electron and its neutrino'',} \textit{ Nucl.
  Phys. B} \textbf{ 90} (1975) 104,
\href{http://dx.doi.org/10.1016/0550-3213(75)90636-7}{\doi{10.1016/0550-3213(75)90636-7}}.

\bibitem{Barbieri:1982eh}
\hrefCMSnoop {}{R.~Barbieri, S.~Ferrara, and C.~A. Savoy, ``Gauge models with
  spontaneously broken local supersymmetry'',} \textit{ Phys. Lett. B} \textbf{
  119} (1982) 343,
\href{http://dx.doi.org/10.1016/0370-2693(82)90685-2}{\doi{10.1016/0370-2693(82)90685-2}}.

\bibitem{Dine:1981rt}
\hrefCMSnoop {}{M.~Dine, W.~Fischler, and M.~Srednicki, ``A simple solution to
  the strong {CP} problem with a harmless axion'',} \textit{ Phys. Lett. B}
  \textbf{ 104} (1981) 199,
\href{http://dx.doi.org/10.1016/0370-2693(81)90590-6}{\doi{10.1016/0370-2693(81)90590-6}}.

\bibitem{Nilles:1982dy}
\hrefCMSnoop {}{H.~P. Nilles, M.~Srednicki, and D.~Wyler, ``Weak interaction
  breakdown induced by supergravity'',} \textit{ Phys. Lett. B} \textbf{ 120}
  (1983) 346,
\href{http://dx.doi.org/10.1016/0370-2693(83)90460-4}{\doi{10.1016/0370-2693(83)90460-4}}.

\bibitem{Frere:1983ag}
\hrefCMSnoop {}{J.~M. Fr{\`e}re, D.~R.~T. Jones, and S.~Raby, ``Fermion masses
  and induction of the weak scale by supergravity'',} \textit{ Nucl. Phys. B}
  \textbf{ 222} (1983) 11,
\href{http://dx.doi.org/10.1016/0550-3213(83)90606-5}{\doi{10.1016/0550-3213(83)90606-5}}.

\bibitem{Derendinger:1983bz}
\hrefCMSnoop {}{J.~P. Derendinger and C.~A. Savoy, ``Quantum effects and {SU(2)
  x U(1)} breaking in supergravity gauge theories'',} \textit{ Nucl. Phys. B}
  \textbf{ 237} (1984) 307,
\href{http://dx.doi.org/10.1016/0550-3213(84)90162-7}{\doi{10.1016/0550-3213(84)90162-7}}.

\bibitem{PhysRevD.39.844}
J.~Ellis\hrefCMSnoop {}{ { et~al.}, ``{Higgs} bosons in a nonminimal
  supersymmetric model'',} \textit{ Phys. Rev. D} \textbf{ 39} (1989) 844,
  \href{http://dx.doi.org/10.1103/PhysRevD.39.844}{\doi{10.1103/PhysRevD.39.844}}.

\bibitem{Drees:1988fc}
\hrefCMSnoop {}{M.~Drees, ``Supersymmetric models with extended {Higgs}
  sector'',} \textit{ Int. J. Mod. Phys. A} \textbf{ 4} (1989) 3635,
\href{http://dx.doi.org/10.1142/S0217751X89001448}{\doi{10.1142/S0217751X89001448}}.

\bibitem{Ellwanger:1993xa}
\hrefCMSnoop {}{U.~Ellwanger, M.~Rausch~de Traubenberg, and C.~A. Savoy,
  ``Particle spectrum in supersymmetric models with a gauge singlet'',}
  \textit{ Phys. Lett. B} \textbf{ 315} (1993) 331,
  \href{http://dx.doi.org/10.1016/0370-2693(93)91621-S}{\doi{10.1016/0370-2693(93)91621-S}},
\href{http://www.arXiv.org/abs/hep-ph/9307322}{\texttt{arXiv:hep-ph/9307322}}.

\bibitem{Ellwanger:1995ru}
\hrefCMSnoop {}{U.~Ellwanger, M.~Rausch~de Traubenberg, and C.~A. Savoy,
  ``{Higgs} phenomenology of the supersymmetric model with a gauge singlet'',}
  \textit{ Z. Phys. C} \textbf{ 67} (1995) 665,
  \href{http://dx.doi.org/10.1007/BF01553993}{\doi{10.1007/BF01553993}},
\href{http://www.arXiv.org/abs/hep-ph/9502206}{\texttt{arXiv:hep-ph/9502206}}.

\bibitem{Ellwanger:1996gw}
\hrefCMSnoop {}{U.~Ellwanger, M.~Rausch~de Traubenberg, and C.~A. Savoy,
  ``Phenomenology of supersymmetric models with a singlet'',} \textit{ Nucl.
  Phys. B} \textbf{ 492} (1997) 21,
  \href{http://dx.doi.org/10.1016/S0550-3213(97)00128-4}{\doi{10.1016/S0550-3213(97)00128-4}},
\href{http://www.arXiv.org/abs/hep-ph/9611251}{\texttt{arXiv:hep-ph/9611251}}.

\bibitem{Elliott:1994ht}
\hrefCMSnoop {}{T.~Elliott, S.~F. King, and P.~L. White, ``Unification
  constraints in the next-to-minimal supersymmetric standard model'',} \textit{
  Phys. Lett. B} \textbf{ 351} (1995) 213,
  \href{http://dx.doi.org/10.1016/0370-2693(95)00381-T}{\doi{10.1016/0370-2693(95)00381-T}},
\href{http://www.arXiv.org/abs/hep-ph/9406303}{\texttt{arXiv:hep-ph/9406303}}.

\bibitem{King:1995vk}
\hrefCMSnoop {}{S.~F. King and P.~L. White, ``Resolving the constrained minimal
  and next-to-minimal supersymmetric standard models'',} \textit{ Phys. Rev. D}
  \textbf{ 52} (1995) 4183,
  \href{http://dx.doi.org/10.1103/PhysRevD.52.4183}{\doi{10.1103/PhysRevD.52.4183}},
\href{http://www.arXiv.org/abs/hep-ph/9505326}{\texttt{arXiv:hep-ph/9505326}}.

\bibitem{Franke:1995tc}
\hrefCMSnoop {}{F.~Franke, ``Neutralinos and {Higgs} bosons in the
  next-to-minimal supersymmetric standard model'',} \textit{ Int. J. Mod. Phys.
  A} \textbf{ 12} (1997) 479,
  \href{http://dx.doi.org/10.1142/S0217751X97000529}{\doi{10.1142/S0217751X97000529}},
\href{http://www.arXiv.org/abs/hep-ph/9512366}{\texttt{arXiv:hep-ph/9512366}}.

\bibitem{Maniatis:2009re}
\hrefCMSnoop {}{M.~Maniatis, ``The next-to-minimal supersymmetric extension of
  the standard model reviewed'',} \textit{ Int. J. Mod. Phys. A} \textbf{ 25}
  (2010) 3505,
  \href{http://dx.doi.org/10.1142/S0217751X10049827}{\doi{10.1142/S0217751X10049827}},
\href{http://www.arXiv.org/abs/0906.0777}{\texttt{arXiv:0906.0777}}.

\bibitem{Ellwanger:2009dp}
\hrefCMSnoop {}{U.~Ellwanger, C.~Hugonie, and A.~M. Teixeira, ``The
  next-to-minimal supersymmetric standard model'',} \textit{ Phys. Rept.}
  \textbf{ 496} (2010) 1,
  \href{http://dx.doi.org/10.1016/j.physrep.2010.07.001}{\doi{10.1016/j.physrep.2010.07.001}},
\href{http://www.arXiv.org/abs/0910.1785}{\texttt{arXiv:0910.1785}}.

\bibitem{1674-1137-38-7-073101}
J.-W. Fan\hrefCMSnoop {}{ { et~al.}, ``Study of diphoton decays of the lightest
  scalar {Higgs} boson in the next-to-minimal supersymmetric standard model'',}
  \textit{ Chin. Phys. C} \textbf{ 38} (2014) 073101,
  \href{http://dx.doi.org/10.1088/1674-1137/38/7/073101}{\doi{10.1088/1674-1137/38/7/073101}},
  \href{http://www.arXiv.org/abs/1309.6394}{\texttt{arXiv:1309.6394}}.

\bibitem{Ellwanger:2015uaz}
\hrefCMSnoop {}{U.~Ellwanger and M.~Rodriguez-Vazquez, ``Discovery prospects of
  a light scalar in the {NMSSM}'',} \textit{ JHEP} \textbf{ 02} (2016) 096,
  \href{http://dx.doi.org/10.1007/JHEP02(2016)096}{\doi{10.1007/JHEP02(2016)096}},
\href{http://www.arXiv.org/abs/1512.04281}{\texttt{arXiv:1512.04281}}.

\bibitem{Guchait:2016pes}
\hrefCMSnoop {}{M.~Guchait and J.~Kumar, ``Diphoton signal of light
  pseudoscalar in {NMSSM} at the {LHC}'',} \textit{ Phys. Rev. D} \textbf{ 95}
  (2017) 035036,
  \href{http://dx.doi.org/10.1103/PhysRevD.95.035036}{\doi{10.1103/PhysRevD.95.035036}},
\href{http://www.arXiv.org/abs/1608.05693}{\texttt{arXiv:1608.05693}}.

\bibitem{Cao:2016uwt}
J.~Cao\hrefCMSnoop {}{ { et~al.}, ``Diphoton signal of the light {Higgs} boson
  in natural {NMSSM}'',} \textit{ Phys. Rev. D} \textbf{ 95} (2017) 116001,
  \href{http://dx.doi.org/10.1103/PhysRevD.95.116001}{\doi{10.1103/PhysRevD.95.116001}},
\href{http://www.arXiv.org/abs/1612.08522}{\texttt{arXiv:1612.08522}}.

\bibitem{Tao_2018}
J.-Q. Tao\hrefCMSnoop {}{ { et~al.}, ``Search for a lighter {Higgs} boson in
  the next-to-minimal supersymmetric standard model'',} \textit{ Chin. Phys. C}
  \textbf{ 42} (2018) 103107,
  \href{http://dx.doi.org/10.1088/1674-1137/42/10/103107}{\doi{10.1088/1674-1137/42/10/103107}},
\href{http://www.arXiv.org/abs/1805.11438}{\texttt{arXiv:1805.11438}}.

\bibitem{GEORGI1985463}
\hrefCMSnoop {}{H.~Georgi and M.~Machacek, ``Doubly-charged {Higgs} bosons'',}
  \textit{ Nucl. Phys. B} \textbf{ 262} (1985) 463,
  \href{http://dx.doi.org/10.1016/0550-3213(85)90325-6}{\doi{10.1016/0550-3213(85)90325-6}}.

\bibitem{PhysRevD.92.075011}
\hrefCMSnoop {}{H.~E. Logan and V.~Rentala, ``All the generalized
  {Georgi-Machacek} models'',} \textit{ Phys. Rev. D} \textbf{ 92} (2015)
  075011,
  \href{http://dx.doi.org/10.1103/PhysRevD.92.075011}{\doi{10.1103/PhysRevD.92.075011}},
\href{http://www.arXiv.org/abs/1502.01275}{\texttt{arXiv:1502.01275}}.

\bibitem{PhysRevD.103.095010}
\hrefCMSnoop {}{A.~Ismail, B.~Keeshan, H.~E. Logan, and Y.~Wu, ``Benchmark for
  {LHC} searches for low-mass custodial fiveplet scalars in the
  {Georgi-Machacek} model'',} \textit{ Phys. Rev. D} \textbf{ 103} (2021)
  095010,
  \href{http://dx.doi.org/10.1103/PhysRevD.103.095010}{\doi{10.1103/PhysRevD.103.095010}},
\href{http://www.arXiv.org/abs/2003.05536}{\texttt{arXiv:2003.05536}}.

\bibitem{Wang_2022}
C.~Wang\hrefCMSnoop {}{ { et~al.}, ``Search for a lighter neutral custodial
  fiveplet scalar in the {Georgi-Machacek} model'',} \textit{ Chin. Phys. C}
  \textbf{ 46} (2022) 083107,
  \href{http://dx.doi.org/10.1088/1674-1137/ac6cd3}{\doi{10.1088/1674-1137/ac6cd3}},
\href{http://www.arXiv.org/abs/2204.09198}{\texttt{arXiv:2204.09198}}.

\bibitem{Barate:2003sz}
\hrefCMSnoop {}{{ALEPH Collaboration, DELPHI Collaboration, L3 Collaboration,
  OPAL Collaboration and the LEP working group for Higgs boson searches},
  ``Search for the standard model {Higgs} boson at {LEP}'',} \textit{ Phys.
  Lett. B} \textbf{ 565} (2003) 61,
  \href{http://dx.doi.org/10.1016/S0370-2693(03)00614-2}{\doi{10.1016/S0370-2693(03)00614-2}},
\href{http://www.arXiv.org/abs/hep-ex/0306033}{\texttt{arXiv:hep-ex/0306033}}.

\bibitem{PhysRevLett.113.171801}
\hrefCMSnoop {}{{ATLAS Collaboration}, ``Search for scalar diphoton resonances
  in the mass range 65--600~{GeV} with the {ATLAS} detector in {$\Pp\Pp$}
  collision data at $\sqrt{s}=8\text{ }\text{ }\mathrm{TeV}$'',} \textit{ Phys.
  Rev. Lett.} \textbf{ 113} (2014) 171801,
  \href{http://dx.doi.org/10.1103/PhysRevLett.113.171801}{\doi{10.1103/PhysRevLett.113.171801}},
  \href{http://www.arXiv.org/abs/1407.6583}{\texttt{arXiv:1407.6583}}.

\bibitem{Aad2023}
\hrefCMSnoop {}{{ATLAS Collaboration}, ``Search for boosted diphoton resonances
  in the 10 to 70~{GeV} mass range using 138\fbinv of 13~{TeV} {$\Pp\Pp$}
  collisions with the {ATLAS} detector'',} \textit{ JHEP} \textbf{ 07} (2023)
  155,
  \href{http://dx.doi.org/10.1007/JHEP07(2023)155}{\doi{10.1007/JHEP07(2023)155}},
  \href{http://www.arXiv.org/abs/2211.04172}{\texttt{arXiv:2211.04172}}.

\bibitem{HIG-17-013}
\hrefCMSnoop {}{{CMS Collaboration}, ``Search for a standard model-like {Higgs}
  boson in the mass range between 70 and 110 {GeV} in the diphoton final state
  in proton-proton collisions at $\sqrt{s}=8$ and $13$ {TeV}'',} \textit{ Phys.
  Lett. B} \textbf{ 793} (2019) 320,
  \href{http://dx.doi.org/10.1016/j.physletb.2019.03.064}{\doi{10.1016/j.physletb.2019.03.064}},
  \href{http://www.arXiv.org/abs/1811.08459}{\texttt{arXiv:1811.08459}}.

\bibitem{hepdata}
\hrefCMSnoop {}{}{HEPD}ata record for this analysis, 2024.
\newblock
  \href{http://dx.doi.org/10.17182/hepdata.150508}{\doi{10.17182/hepdata.150508}}.

\bibitem{HIG-13-001}
\hrefCMSnoop {}{{CMS Collaboration}, ``Observation of the diphoton decay of the
  {Higgs} boson and measurement of its properties'',} \textit{ Eur. Phys. J. C}
  \textbf{ 74} (2014) 3076,
  \href{http://dx.doi.org/10.1140/epjc/s10052-014-3076-z}{\doi{10.1140/epjc/s10052-014-3076-z}},
\href{http://www.arXiv.org/abs/1407.0558}{\texttt{arXiv:1407.0558}}.

\bibitem{CMS-PAS-HIG-16-040}
\hrefCMSnoop {}{{CMS Collaboration}, ``Measurements of {Higgs} boson properties
  in the diphoton decay channel in proton-proton collisions at $\sqrt{s} =$ 13
  {TeV}'',} \textit{ JHEP} \textbf{ 11} (2018) 185,
  \href{http://dx.doi.org/10.1007/JHEP11(2018)185}{\doi{10.1007/JHEP11(2018)185}},
\href{http://www.arXiv.org/abs/1804.02716}{\texttt{arXiv:1804.02716}}.

\bibitem{Chatrchyan:2008zzk}
\hrefCMSnoop {}{{CMS Collaboration}, ``The {CMS} experiment at the {CERN
  LHC}'',} \textit{ JINST} \textbf{ 3} (2008) S08004,
\href{http://dx.doi.org/10.1088/1748-0221/3/08/S08004}{\doi{10.1088/1748-0221/3/08/S08004}}.

\bibitem{CMS:EGM-14-001}
\hrefCMSnoop {}{{CMS Collaboration}, ``Performance of photon reconstruction and
  identification with the {CMS} detector in proton-proton collisions at
  $\sqrt{s}=8$ {TeV}'',} \textit{ JINST} \textbf{ 10} (2015) P08010,
  \href{http://dx.doi.org/10.1088/1748-0221/10/08/P08010}{\doi{10.1088/1748-0221/10/08/P08010}},
\href{http://www.arXiv.org/abs/1502.02702}{\texttt{arXiv:1502.02702}}.

\bibitem{CMS:2020xrn}
\hrefCMSnoop {}{{CMS Collaboration}, ``A measurement of the {Higgs} boson mass
  in the diphoton decay channel'',} \textit{ Phys. Lett. B} \textbf{ 805}
  (2020) 135425,
  \href{http://dx.doi.org/10.1016/j.physletb.2020.135425}{\doi{10.1016/j.physletb.2020.135425}},
  \href{http://www.arXiv.org/abs/2002.06398}{\texttt{arXiv:2002.06398}}.

\bibitem{CMS-PRF-14-001}
\hrefCMSnoop {}{{CMS Collaboration}, ``Particle-flow reconstruction and global
  event description with the {CMS} detector'',} \textit{ JINST} \textbf{ 12}
  (2017) P10003,
  \href{http://dx.doi.org/10.1088/1748-0221/12/10/P10003}{\doi{10.1088/1748-0221/12/10/P10003}},
\href{http://www.arXiv.org/abs/1706.04965}{\texttt{arXiv:1706.04965}}.

\bibitem{Alwall:2014hca}
J.~Alwall\hrefCMSnoop {}{ { et~al.}, ``The automated computation of tree-level
  and next-to-leading order differential cross sections, and their matching to
  parton shower simulations'',} \textit{ JHEP} \textbf{ 07} (2014) 079,
  \href{http://dx.doi.org/10.1007/JHEP07(2014)079}{\doi{10.1007/JHEP07(2014)079}},
\href{http://www.arXiv.org/abs/1405.0301}{\texttt{arXiv:1405.0301}}.

\bibitem{FXFX}
\hrefCMSnoop {}{R.~Frederix and S.~Frixione, ``Merging meets matching in
  {MC@NLO}'',} \textit{ JHEP} \textbf{ 12} (2012) 061,
  \href{http://dx.doi.org/10.1007/JHEP12(2012)061}{\doi{10.1007/JHEP12(2012)061}},
\href{http://www.arXiv.org/abs/1209.6215}{\texttt{arXiv:1209.6215}}.

\bibitem{Ball:2014uwa}
\hrefCMSnoop {}{{NNPDF} Collaboration, ``Parton distributions for the {LHC} run
  {II}'',} \textit{ JHEP} \textbf{ 04} (2015) 040,
  \href{http://dx.doi.org/10.1007/JHEP04(2015)040}{\doi{10.1007/JHEP04(2015)040}},
\href{http://www.arXiv.org/abs/1410.8849}{\texttt{arXiv:1410.8849}}.

\bibitem{NNPDF31}
\hrefCMSnoop {}{{NNPDF} Collaboration, ``{Parton distributions from
  high-precision collider data}'',} \textit{ Eur. Phys. J. C} \textbf{ 77}
  (2017) 663,
  \href{http://dx.doi.org/10.1140/epjc/s10052-017-5199-5}{\doi{10.1140/epjc/s10052-017-5199-5}},
  \href{http://www.arXiv.org/abs/1706.00428}{\texttt{arXiv:1706.00428}}.

\bibitem{Hamilton:2013fea}
\hrefCMSnoop {}{K.~Hamilton, P.~Nason, E.~Re, and G.~Zanderighi, ``{NNLOPS
  simulation of Higgs boson production}'',} \textit{ JHEP} \textbf{ 10} (2013)
  222,
  \href{http://dx.doi.org/10.1007/JHEP10(2013)222}{\doi{10.1007/JHEP10(2013)222}},
\href{http://www.arXiv.org/abs/1309.0017}{\texttt{arXiv:1309.0017}}.

\bibitem{Sjostrand:2014zea}
T.~Sj{\"o}strand\hrefCMSnoop {}{ { et~al.}, ``{An introduction to PYTHIA
  8.2}'',} \textit{ Comput. Phys. Commun.} \textbf{ 191} (2015) 159,
  \href{http://dx.doi.org/10.1016/j.cpc.2015.01.024}{\doi{10.1016/j.cpc.2015.01.024}},
\href{http://www.arXiv.org/abs/1410.3012}{\texttt{arXiv:1410.3012}}.

\bibitem{Khachatryan:2015pea}
\hrefCMSnoop {}{{CMS Collaboration}, ``Event generator tunes obtained from
  underlying event and multiparton scattering measurements'',} \textit{ Eur.
  Phys. J. C} \textbf{ 76} (2016) 155,
  \href{http://dx.doi.org/10.1140/epjc/s10052-016-3988-x}{\doi{10.1140/epjc/s10052-016-3988-x}},
\href{http://www.arXiv.org/abs/1512.00815}{\texttt{arXiv:1512.00815}}.

\bibitem{Sirunyan:2019dfx}
\hrefCMSnoop {}{{CMS Collaboration}, ``{Extraction and validation of a new set
  of CMS PYTHIA8 tunes from underlying-event measurements}'',} \textit{ Eur.
  Phys. J. C} \textbf{ 80} (2020) 4,
  \href{http://dx.doi.org/10.1140/epjc/s10052-019-7499-4}{\doi{10.1140/epjc/s10052-019-7499-4}},
\href{http://www.arXiv.org/abs/1903.12179}{\texttt{arXiv:1903.12179}}.

\bibitem{deFlorian:2016spz}
\hrefCMSnoop {}{{LHC Higgs Cross Section Working Group}, ``Handbook of {LHC}
  {Higgs} cross sections: 4. deciphering the nature of the {Higgs} sector'',}
  \textit{ CERN} (2016)
  \href{http://dx.doi.org/10.23731/CYRM-2017-002}{\doi{10.23731/CYRM-2017-002}},
\href{http://www.arXiv.org/abs/1610.07922}{\texttt{arXiv:1610.07922}}.

\bibitem{GEANT}
\hrefCMSnoop {}{{GEANT4} Collaboration, ``{\GEANTfour}---a simulation
  toolkit'',} \textit{ Nucl. Instrum. Meth. A} \textbf{ 506} (2003) 250,
\href{http://dx.doi.org/10.1016/S0168-9002(03)01368-8}{\doi{10.1016/S0168-9002(03)01368-8}}.

\bibitem{Gleisberg:2008ta}
T.~Gleisberg\hrefCMSnoop {}{ { et~al.}, ``Event generation with {SHERPA}
  1.1'',} \textit{ JHEP} \textbf{ 02} (2009) 007,
  \href{http://dx.doi.org/10.1088/1126-6708/2009/02/007}{\doi{10.1088/1126-6708/2009/02/007}},
\href{http://www.arXiv.org/abs/0811.4622}{\texttt{arXiv:0811.4622}}.

\bibitem{CMS:2020cmk}
\hrefCMSnoop {}{{CMS Collaboration}, ``{Performance of the CMS Level-1 trigger
  in proton-proton collisions at $\sqrt{s} = 13$\,TeV}'',} \textit{ JINST}
  \textbf{ 15} (2020) P10017,
  \href{http://dx.doi.org/10.1088/1748-0221/15/10/P10017}{\doi{10.1088/1748-0221/15/10/P10017}},
  \href{http://www.arXiv.org/abs/2006.10165}{\texttt{arXiv:2006.10165}}.

\bibitem{Khachatryan:2016bia}
\hrefCMSnoop {}{{CMS Collaboration}, ``The {CMS} trigger system'',} \textit{
  JINST} \textbf{ 12} (2017) P01020,
  \href{http://dx.doi.org/10.1088/1748-0221/12/01/P01020}{\doi{10.1088/1748-0221/12/01/P01020}},
\href{http://www.arXiv.org/abs/1609.02366}{\texttt{arXiv:1609.02366}}.

\bibitem{Sirunyan:2020ycc}
\hrefCMSnoop {}{{CMS Collaboration}, ``{Electron and photon reconstruction and
  identification with the CMS experiment at the CERN LHC}'',} \textit{ JINST}
  \textbf{ 16} (2021) P05014,
  \href{http://dx.doi.org/10.1088/1748-0221/16/05/P05014}{\doi{10.1088/1748-0221/16/05/P05014}},
  \href{http://www.arXiv.org/abs/2012.06888}{\texttt{arXiv:2012.06888}}.

\bibitem{tagNprobe}
\hrefCMSnoop {}{{CMS Collaboration}, ``Measurement of the inclusive {\PW} and
  {\PZ} production cross sections in {$\Pp\Pp$} collisions at $\sqrt{s}=7$
  {TeV}'',} \textit{ JHEP} \textbf{ 10} (2011) 132,
  \href{http://dx.doi.org/10.1007/JHEP10(2011)132}{\doi{10.1007/JHEP10(2011)132}},
\href{http://www.arXiv.org/abs/1107.4789}{\texttt{arXiv:1107.4789}}.

\bibitem{Chatrchyan:2013dga}
\hrefCMSnoop {}{{CMS Collaboration}, ``Energy calibration and resolution of the
  {CMS} electromagnetic calorimeter in {$\Pp\Pp$} collisions at $\sqrt{s} = 7$
  {TeV}'',} \textit{ JINST} \textbf{ 8} (2013) P09009,
  \href{http://dx.doi.org/10.1088/1748-0221/8/09/P09009}{\doi{10.1088/1748-0221/8/09/P09009}},
\href{http://www.arXiv.org/abs/1306.2016}{\texttt{arXiv:1306.2016}}.

\bibitem{Cacciari:2008gp}
\hrefCMSnoop {}{M.~Cacciari, G.~P. Salam, and G.~Soyez, ``{The anti-\kt jet
  clustering algorithm}'',} \textit{ JHEP} \textbf{ 04} (2008) 063,
  \href{http://dx.doi.org/10.1088/1126-6708/2008/04/063}{\doi{10.1088/1126-6708/2008/04/063}},
  \href{http://www.arXiv.org/abs/0802.1189}{\texttt{arXiv:0802.1189}}.

\bibitem{Cacciari:2011ma}
\hrefCMSnoop {}{M.~Cacciari, G.~P. Salam, and G.~Soyez, ``{FastJet user
  manual}'',} \textit{ Eur. Phys. J. C} \textbf{ 72} (2012) 1896,
  \href{http://dx.doi.org/10.1140/epjc/s10052-012-1896-2}{\doi{10.1140/epjc/s10052-012-1896-2}},
\href{http://www.arXiv.org/abs/1111.6097}{\texttt{arXiv:1111.6097}}.

\bibitem{CMS:2020ebo}
\hrefCMSnoop {}{{CMS Collaboration}, ``{Pileup mitigation at CMS in 13\TeV
  data}'',} \textit{ JINST} \textbf{ 15} (2020) P09018,
  \href{http://dx.doi.org/10.1088/1748-0221/15/09/P09018}{\doi{10.1088/1748-0221/15/09/P09018}},
  \href{http://www.arXiv.org/abs/2003.00503}{\texttt{arXiv:2003.00503}}.

\bibitem{Khachatryan:2016kdb}
\hrefCMSnoop {}{{CMS Collaboration}, ``Jet energy scale and resolution in the
  {CMS} experiment in {$\Pp\Pp$} collisions at 8 {TeV}'',} \textit{ JINST}
  \textbf{ 12} (2017) P02014,
  \href{http://dx.doi.org/10.1088/1748-0221/12/02/P02014}{\doi{10.1088/1748-0221/12/02/P02014}},
\href{http://www.arXiv.org/abs/1607.03663}{\texttt{arXiv:1607.03663}}.

\bibitem{tmva}
\hrefCMSnoop {}{A.~Hoecker { et~al.}, ``{TMVA}---toolkit for multivariate data
  analysis'',} 2007.
\href{http://www.arXiv.org/abs/physics/0703039}{\texttt{arXiv:physics/0703039}}.

\bibitem{Pedregosa:2011ork}
\hrefCMSnoop {}{F.~Pedregosa { et~al.}, ``Scikit-learn: machine learning in
  {Python}'',} \textit{ J. Machine Learning Res.} \textbf{ 12} (2011) 2825,
  \href{http://dx.doi.org/10.48550/arXiv.1201.0490}{\doi{10.48550/arXiv.1201.0490}},
\href{http://www.arXiv.org/abs/1201.0490}{\texttt{arXiv:1201.0490}}.

\bibitem{ftest}
\hrefCMSnoop {}{R.~A. Fisher, ``On the interpretation of $\chi^{2}$ from
  contingency tables, and the calculation of~$p$'',} \textit{ J. Royal Stat.
  Soc} \textbf{ 85} (1922) 87,
  \href{http://dx.doi.org/10.2307/2340521}{\doi{10.2307/2340521}}.

\bibitem{DiscreteProfilingMethod}
\hrefCMSnoop {}{P.~D. Dauncey, M.~Kenzie, N.~Wardle, and G.~J. Davies,
  ``Handling uncertainties in background shapes: the discrete profiling
  method'',} \textit{ JINST} \textbf{ 10} (2015) P04015,
  \href{http://dx.doi.org/10.1088/1748-0221/10/04/P04015}{\doi{10.1088/1748-0221/10/04/P04015}},
\href{http://www.arXiv.org/abs/1408.6865}{\texttt{arXiv:1408.6865}}.

\bibitem{CrystalBall}
\href
  {http://www.slac.stanford.edu/pubs/slacreports/slac-r-236.html}{M.~Oreglia,
  ``A study of the reactions $\psi^\prime \to \gamma \gamma \psi$''}.
\newblock PhD thesis, Stanford University, 1980.
\newblock {{SLAC} Report {SLAC-R-236}}.

\bibitem{DoubleCrystalBallRef}
\href
  {http://www-library.desy.de/preparch/desy/int_rep/f31-86-02.pdf}{T.~Skwarnicki,
  ``A study of the radiative {CASCADE} transitions between the {Upsilon}-prime
  and {Upsilon} resonances''}.
\newblock PhD thesis, Cracow, INP, 1986.

\bibitem{CMS-LUM-17-003}
\hrefCMSnoop {}{{CMS Collaboration}, ``Precision luminosity measurement in
  proton-proton collisions at $\sqrt{s} =$ 13 {TeV} in 2015 and 2016 at
  {CMS}'',} \textit{ Eur. Phys. J. C} \textbf{ 81} (2021) 800,
  \href{http://dx.doi.org/10.1140/epjc/s10052-021-09538-2}{\doi{10.1140/epjc/s10052-021-09538-2}},
  \href{http://www.arXiv.org/abs/2104.01927}{\texttt{arXiv:2104.01927}}.

\bibitem{CMSlumi2017}
\href {https://cds.cern.ch/record/2621960}{{CMS Collaboration}, ``{CMS
  luminosity measurement for the 2017 data-taking period at $\sqrt{s} =
  13~\mathrm{TeV}$}'',} CMS Physics Analysis Summary CMS-PAS-LUM-17-004, 2018.

\bibitem{CMSlumi2018}
\href {https://cds.cern.ch/record/2676164}{{CMS Collaboration}, ``{CMS
  luminosity measurement for the 2018 data-taking period at $\sqrt{s} =
  13~\mathrm{TeV}$}'',} CMS Physics Analysis Summary CMS-PAS-LUM-18-002, 2019.

\bibitem{Carrazza2015}
S.~Carrazza\hrefCMSnoop {}{ { et~al.}, ``An unbiased {Hessian} representation
  for {Monte Carlo} {PDFs}'',} \textit{ Eur. Phys. J. C} \textbf{ 75} (2015)
  369,
  \href{http://dx.doi.org/10.1140/epjc/s10052-015-3590-7}{\doi{10.1140/epjc/s10052-015-3590-7}},
\href{http://www.arXiv.org/abs/1505.06736}{\texttt{arXiv:1505.06736}}.

\bibitem{Gao_2014}
\hrefCMSnoop {}{J.~Gao and P.~Nadolsky, ``{A meta-analysis of parton
  distribution functions}'',} \textit{ JHEP} \textbf{ 07} (2014) 035,
  \href{http://dx.doi.org/10.1007/jhep07(2014)035}{\doi{10.1007/jhep07(2014)035}},
  \href{http://www.arXiv.org/abs/1401.0013}{\texttt{arXiv:1401.0013}}.

\bibitem{Butterworth:2015oua}
\hrefCMSnoop {}{J.~Butterworth { et~al.}, ``{PDF4LHC} recommendations for {LHC
  Run II}'',} \textit{ J. Phys. G} \textbf{ 43} (2016) 023001,
  \href{http://dx.doi.org/10.1088/0954-3899/43/2/023001}{\doi{10.1088/0954-3899/43/2/023001}},
\href{http://www.arXiv.org/abs/1510.03865}{\texttt{arXiv:1510.03865}}.

\bibitem{HIG-19-015}
\hrefCMSnoop {}{{CMS Collaboration}, ``Measurements of {Higgs} boson production
  cross sections and couplings in the diphoton decay channel at
  $\sqrt{\mathrm{s}} =$ 13 {TeV}'',} \textit{ JHEP} \textbf{ 07} (2021) 027,
  \href{http://dx.doi.org/10.1007/jhep07(2021)027}{\doi{10.1007/jhep07(2021)027}},
\href{http://www.arXiv.org/abs/2103.06956}{\texttt{arXiv:2103.06956}}.

\bibitem{LHC-HCG}
\href {http://cdsweb.cern.ch/record/1379837}{{ATLAS and CMS Collaborations, LHC
  Higgs Combination Group}, ``Procedure for the {LHC} {Higgs} boson search
  combination in {Summer} 2011'',} CMS Note CMS-NOTE-2011-005,
  ATL-PHYS-PUB-2011-11, CERN, 2011.

\bibitem{Junk:1999kv}
\hrefCMSnoop {}{T.~Junk, ``Confidence level computation for combining searches
  with small statistics'',} \textit{ Nucl. Instrum. Meth. A} \textbf{ 434}
  (1999) 435,
  \href{http://dx.doi.org/10.1016/S0168-9002(99)00498-2}{\doi{10.1016/S0168-9002(99)00498-2}},
\href{http://www.arXiv.org/abs/hep-ex/9902006}{\texttt{arXiv:hep-ex/9902006}}.

\bibitem{Read:2000ru}
\hrefCMSnoop {}{A.~L. Read, ``Presentation of search results: the {\CLs}
  technique'',} \textit{ J. Phys. G} \textbf{ 28} (2002) 2693,
\href{http://dx.doi.org/10.1088/0954-3899/28/10/313}{\doi{10.1088/0954-3899/28/10/313}}.

\bibitem{Cowan:2010js}
\hrefCMSnoop {}{G.~Cowan, K.~Cranmer, E.~Gross, and O.~Vitells, ``Asymptotic
  formulae for likelihood-based tests of new physics'',} \textit{ Eur. Phys. J.
  C} \textbf{ 71} (2011) 1554,
  \href{http://dx.doi.org/10.1140/epjc/s10052-011-1554-0}{\doi{10.1140/epjc/s10052-011-1554-0}},
  \href{http://www.arXiv.org/abs/1007.1727}{\texttt{arXiv:1007.1727}}.
[Erratum: \DOI{10.1140/epjc/s10052-013-2501-z}].

\bibitem{pvalue}
\hrefCMSnoop {}{L.~Demortier, ``{P}-values and nuisance parameters'',} in
  \textit{ Statistical issues for {LHC} physics. {Proceedings, Workshop,
  PHYSTAT-LHC, Geneva, Switzerland, June} 27-29, 2007}, p.~23.
\newblock 2008.
\newblock
\href{http://dx.doi.org/10.5170/CERN-2008-001}{\doi{10.5170/CERN-2008-001}}.

\end{thebibliography}\endgroup

\cleardoublepage \appendix\section{The CMS Collaboration \label{app:collab}}\begin{sloppypar}\hyphenpenalty=5000\widowpenalty=500\clubpenalty=5000
\cmsinstitute{Yerevan Physics Institute, Yerevan, Armenia}
{\tolerance=6000
A.~Hayrapetyan, A.~Tumasyan\cmsAuthorMark{1}\cmsorcid{0009-0000-0684-6742}
\par}
\cmsinstitute{Institut f\"{u}r Hochenergiephysik, Vienna, Austria}
{\tolerance=6000
W.~Adam\cmsorcid{0000-0001-9099-4341}, J.W.~Andrejkovic, T.~Bergauer\cmsorcid{0000-0002-5786-0293}, S.~Chatterjee\cmsorcid{0000-0003-2660-0349}, K.~Damanakis\cmsorcid{0000-0001-5389-2872}, M.~Dragicevic\cmsorcid{0000-0003-1967-6783}, A.~Escalante~Del~Valle\cmsorcid{0000-0002-9702-6359}, P.S.~Hussain\cmsorcid{0000-0002-4825-5278}, M.~Jeitler\cmsAuthorMark{2}\cmsorcid{0000-0002-5141-9560}, N.~Krammer\cmsorcid{0000-0002-0548-0985}, A.~Li\cmsorcid{0000-0002-4547-116X}, D.~Liko\cmsorcid{0000-0002-3380-473X}, I.~Mikulec\cmsorcid{0000-0003-0385-2746}, J.~Schieck\cmsAuthorMark{2}\cmsorcid{0000-0002-1058-8093}, R.~Sch\"{o}fbeck\cmsorcid{0000-0002-2332-8784}, D.~Schwarz\cmsorcid{0000-0002-3821-7331}, M.~Sonawane\cmsorcid{0000-0003-0510-7010}, S.~Templ\cmsorcid{0000-0003-3137-5692}, W.~Waltenberger\cmsorcid{0000-0002-6215-7228}, C.-E.~Wulz\cmsAuthorMark{2}\cmsorcid{0000-0001-9226-5812}
% ...
\par}
\vskip\cmsinstskip
\dag:~Deceased\\
$^{1}$Also at Yerevan State University, Yerevan, Armenia\\
$^{2}$Also at TU Wien, Vienna, Austria\\
% ...
\end{sloppypar}
\end{document}